\RequirePackage{fix-cm} 
\documentclass[a4paper, twoside, reqno, 12pt, dvips]{amsart}
\usepackage{fixltx2e}     

\usepackage[english]{babel}
\usepackage[latin1]{inputenc}

\usepackage{eucal}
\usepackage{esint}
\usepackage{amsgen}
\usepackage{amsthm}
\usepackage{amssymb}
\usepackage{amsmath}
\usepackage{amsfonts}
\usepackage{verbatim}
\usepackage[nice]{nicefrac}
\usepackage{mathrsfs, dsfont}

\usepackage{a4wide}
\usepackage{xspace}

\usepackage{indentfirst}
\usepackage{graphicx}
\usepackage{subfigure}
\usepackage[section]{placeins}
\usepackage{psfrag}
\usepackage{infix-RPN}

\usepackage{microtype}
\usepackage[bf, up, hang]{caption}

\usepackage[usenames, dvipsnames, pdf]{pstricks}
\usepackage{epsfig}
\usepackage{pst-grad} 
\usepackage{pst-plot} 

\usepackage{color}
\definecolor{oneblue}{rgb}{0.0, 0.0, 0.85}
\definecolor{darkgrey}{rgb}{0.273, 0.281, 0.30}

\usepackage[colorlinks,
            urlcolor=oneblue,
            linkcolor=oneblue,
            citecolor=oneblue,
            bookmarksopen=false,
            pagebackref]{hyperref}

\usepackage[compact]{titlesec}

\titleformat{\section}{\normalfont\Large\bfseries\sffamily\center\color{darkgrey}}{\thesection.}{0.5em}{}{}
\titleformat{\subsection}{\normalfont\large\bfseries\sffamily\color{darkgrey}}{\thesubsection.}{0.4em}{}{}
\titleformat{\subsubsection}{\normalfont\normalsize\bfseries\sffamily\color{darkgrey}}{\thesubsubsection.}{0.3em}{}{}

\titlespacing*{\section}{1.0em}{1.0em}{0.8em}[0em]
\titlespacing*{\subsection}{1.0em}{1.0em}{0.8em}[0em]
\titlespacing*{\subsubsection}{1.0em}{0.7em}{0.6em}[0em]

\usepackage{titletoc}

\contentsmargin{0.0em}
\dottedcontents{section}[2.5em]{\addvspace{0.45em}\bfseries}{1.0em}{0pc}
\dottedcontents{subsection}[4.5em]{}{2.6em}{0pc}
\dottedcontents{subsubsection}[5.5em]{}{3.0em}{0pc}

\usepackage{fancyhdr}
\usepackage{lastpage}

\newcommand*\Title{Run-up amplification of long waves}
\newcommand*\Authors{T.~Stefanakis, S.~Xu, D.~Dutykh \& F.~Dias}

\numberwithin{equation}{section}

\newcommand{\m}{{\sf m}}
\newcommand{\s}{{\sf s}}
\newcommand{\ui}{\mathrm{i}}
\renewcommand{\d}{\mathrm{d}}

\begin{document}

\title[\Title]{Run-up amplification of transient long waves}

\author[T.~Stefanakis]{Themistoklis S. Stefanakis}
\address{CMLA, ENS Cachan, CNRS, 61 Avenue du Pr\'esident Wilson, F-94230 Cachan, France \and University College Dublin, School of Mathematical Sciences, Belfield, Dublin 4, Ireland}
\email{themistoklisstef@yahoo.co.uk}

\author[S.~Xu]{Shanshan Xu}
\address{University College Dublin, School of Mathematical Sciences, Belfield, Dublin 4, Ireland}
\email{shanshan.xu@ucdconnect.ie}

\author[D.~Dutykh]{Denys Dutykh}
\address{University College Dublin, School of Mathematical Sciences, Belfield, Dublin 4, Ireland \and LAMA, UMR 5127 CNRS, Universit\'e de Savoie, Campus Scientifique, 73376 Le Bourget-du-Lac Cedex, France}
\email{Denys.Dutykh@ucd.ie}
\urladdr{http://www.denys-dutykh.com/}

\author[F.~Dias]{Fr\'ed\'eric Dias}
\address{CMLA, ENS Cachan, CNRS, 61 Avenue du Pr\'esident Wilson, F-94230 Cachan, France \and University College Dublin, School of Mathematical Sciences, Belfield, Dublin 4, Ireland}
\email{Frederic.Dias@ucd.ie}

\begin{abstract}
The extreme characteristics of the run-up of transient long waves are studied in this paper. First we give a brief overview of the existing theory which is mainly based on the hodograph transformation \cite{CG58}. Then, using numerical simulations, we build on the work of \textsc{Stefanakis} \emph{et al.} (2011) \cite{Stefanakis2011} for an infinite sloping beach and we find that resonant run-up amplification of monochromatic waves is robust to spectral perturbations of the incoming wave and resonant regimes do exist for certain values of the frequency. In the setting of a finite beach attached to a constant depth region, resonance can only be observed when the incoming wavelength is larger than the distance from the undisturbed shoreline to the seaward boundary. Wavefront steepness is also found to play a role in wave run-up, with steeper waves reaching higher run-up values.

\bigskip
\noindent \textbf{\keywordsname:} run-up; long waves; resonance; sloping beach; tsunami
\end{abstract}

\maketitle
\tableofcontents
\thispagestyle{empty}

\section{Introduction}

Long wave run-up, the maximum elevation of wave uprush on a beach above still water level, is difficult to observe in nature in real time due to the large physical dimensions of the phenomenon and to the catastrophic consequences it usually leads to, since the most famous representation of a long wave is that of a tsunami. Tidal waves, meteotsunamis \cite{Rabinovich2009} and storm surges are also long waves.

Most observational data concerning run-up are collected during post-tsunami surveys. Nevertheless, this data does not offer any information, by itself,  on the time history of the event, which leads field scientists to rely on interviews with eye-witnesses, who, in some cases,  have reported that it is not always the first wave which results in the worst damage. Moreover, unexpected extreme localized run-up values have been measured during several tsunami events, such as in Java 1994 \cite{Tsuji1995}, Java 2006 \cite{Fritz2007}, Chile 2010 \cite{Fritz2011} and Japan 2011 \cite{Grilli2012}. Hence, a question rises whether these extreme run-up values are related to non-leading waves.

\textsc{Stefanakis} \emph{et al.} (2011) \cite{Stefanakis2011} showed that for a given plane beach slope there exist wave frequencies that lead to resonant long wave run-up amplification by non-leading waves. These results were confirmed experimentally in a wave tank by \cite{Ezersky2012} who distinguished the frequency that leads to resonant run-up from the resonant frequency of the wavemaker. They also observed a secondary resonant regime which was not identified before. The authors also recognized that the resonant state occurs when the Bessel function $J_0(\sqrt{4\omega^2 L / g \tan \theta}) = 0$, as predicted by the linear theory \cite{Lamb1932}, where $\omega$ is the angular frequency of the wave, $\tan{\theta}$ is the beach slope, $L$ is the horizontal distance from the undisturbed shoreline to the point where the wave amplitude is imposed and $g$ is the gravitational acceleration. Several other possible explanations for the observed extreme run-up values are also available.

\textsc{Miles} (1971) \cite{Miles1971} described the conditions for harbor resonance and the importance of the Helmholtz mode to tsunami response and later \textsc{Kajiura} (1977) \cite{Kajiura1977} introduced the notion of bay resonance. \textsc{Agnon} \& \textsc{Mei} (1988) \cite{Agnon1988} and \textsc{Grataloup} \& \textsc{Mei} (2003) \cite{Grataloup2003} studied the long wave resonance due to wave trapping and wave-wave interactions. \textsc{Munk} \emph{et al.} (1964) \cite{Munk1964} and \textsc{Rabinovich} \& \textsc{Leviant} (1992) \cite{Rabinovich1992} studied wave resonance in the context of shelf resonance, which occurs when tidal waves have a wavelength four times larger than the continental shelf width. \textsc{Fritz} \emph{et al.} (2007) \cite{Fritz2007} suggested that the extreme run-up values measured after the Java 2006 tsunami could be explained by a submarine landslide triggered by the earthquake. All of the above underline the critical role of bathymetry and coastal geometry to long wave propagation and run-up. In a recent study, \textsc{Kanoglu} \emph{et al.} (2013) \cite{Kanoglu2013} argued that finite-crest length effects may produce focusing. Nonetheless, resonant run-up has already been documented for the case of short waves \cite{Bruun1977} with an interesting description: \begin{quote} ``\textit{[Resonant run-up] occurs when run-down is in a low position and wave breaking takes place simultaneously and repeatedly close to that location}." \end{quote} Similar observations have been made by \textsc{Stefanakis} \emph{et al.} (2011) \cite{Stefanakis2011}.

On a theoretical basis, the main mathematical difficulty of the run-up problem is the moving shoreline. Progress was made through the introduction of the Carrier and Greenspan (CG) transformation \cite{CG58} which leads to the reduction of the two Nonlinear Shallow Water Equations (NSWE) into one linear, but the ingenuity of this transformation is that in the transformed space the moving shoreline is static. With the aid of the CG transformation several significant contributions were made to the long wave run-up problem \cite{Keller1964, Carrier1966, Synolakis1987, TS94, Brocchini1996, Kanoglu2006, Antuono2007}.

A thorough review of the long wave run-up problem with additional results on its relation to the surf-similarity is given by \textsc{Madsen} \& \textsc{Fuhrman} (2008) \cite{Madsen2008}. The aforementioned theoretical results do not exhibit any resonant regimes and were reproduced numerically by \textsc{Madsen} \& \textsc{Fuhrman} (2008) \cite{Madsen2008} by placing a relaxation zone close to the wave generation region, which absorbs the reflected wavefield. These sponge layers are widely used because the combination of incoming and outgoing waves at the boundary remains still poorly understood. These are artifacts and are not part of the governing wave equations.

In the present paper we first provide an overview of the theory behind long wave run-up on a plane beach\footnote{Since the theory has been developed over the last five decades, it is useful to provide a short review of the major advances in a condensed form. That way, the alternative approach proposed in the present paper will appear more clearly.} and we confirm the resonance results of \textsc{Stefanakis} \emph{et al.} (2011) \cite{Stefanakis2011} with more geophysically relevant bottom slopes. We also prove their robustness to modal perturbations. The case of a piecewise linear beach follows where we show both analytically and computationally the existence of resonant states. Then we explore whether resonance can be observed  when a sloping beach is connected to a constant depth region and we test the effect of wave nonlinearity and how the results relate to the theory. Finally, we discuss the effect the boundary condition has on the resonant run-up amplification.

\section{Statement of problem and method of analytical solution}\label{sec2}

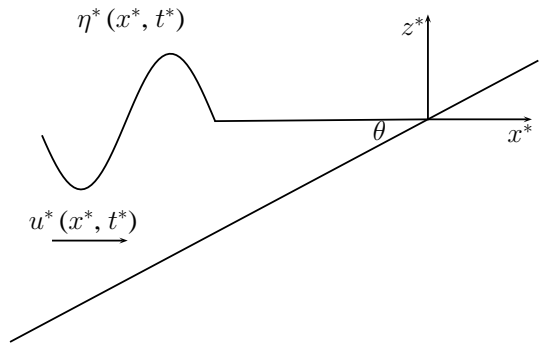
\begin{figure}
\begin{center}
\scalebox{0.6} 
{
\begin{pspicture}(0,-3.64)(11.6,3.64)
\psline[linewidth=0.04cm](0.0,-3.62)(11.58,2.6)
\psline[linewidth=0.04cm](4.48,1.26)(9.18,1.3)
\infixtoRPN{-1.5*sin(1.6*(x-4.5)*180/3.1415)+1.25}
\psplot[linewidth=0.04]
     {0.7}{4.5}{\RPN}
\psline[linewidth=0.04cm,arrowsize=0.05291667cm 2.0,arrowlength=1.4,arrowinset=0.4]{->}(9.16,1.3)(9.16,3.62)
\psline[linewidth=0.04cm,arrowsize=0.05291667cm 2.0,arrowlength=1.4,arrowinset=0.4]{->}(9.16,1.3)(11.44,1.3)
\usefont{T1}{ptm}{m}{n}
\rput(8.83,3.365){\Large $z^*$}
\usefont{T1}{ptm}{m}{n}
\rput(11.19,1.0){\Large $x^*$}
\usefont{T1}{ptm}{m}{n}
\rput(8.1,1.05){\Large $\theta$}
\psline[linewidth=0.04cm,arrowsize=0.05291667cm 2.0,arrowlength=1.4,arrowinset=0.4]{->}(0.92,-1.38)(2.58,-1.38)
\usefont{T1}{ptm}{m}{n}
\rput(1.6,-0.975){\Large $u^*$\,($x^*$, $t^*$)}
\usefont{T1}{ptm}{m}{n}
\rput(2.65,3.5){\Large $\eta^*$\,($x^*$, $t^*$)}
\end{pspicture} 
}
\end{center}
\caption{\em The geometry of the run-up problem along a sloping beach.}
\label{fig:geometry}
\end{figure}

In the following we present a review of the analytical solution. Consider a propagation problem described by the one-dimensional NSWE 
\begin{equation}\label{Riemann1} 
\frac{\partial\eta^*}{\partial t^*}+ \frac{\partial}{\partial x^*}\,[(h+\eta^*)\,u^*]=0,                               
\quad
\frac{\partial u^*}{\partial t^*} + u^* \frac{\partial u^*}{\partial x^*} +
g \frac{\partial\eta^*}{ \partial x^*} = 0
\end{equation}
where $z^*=\eta^*(x^*,t^*)$ is the free surface elevation, $h(x^*)$ is the water depth, $u^*(x^*,t^*)$ is the depth-averaged horizontal velocity and $g$ is the acceleration due to gravity. Consider a topography consisting of a sloping beach with unperturbed water depth varying linearly with the horizontal coordinate, $h(x^*) = - \alpha x^*$, where $\alpha = \tan \theta$ is the bottom slope (see Fig.~\ref{fig:geometry}).

In order to solve equations (\ref{Riemann1}), appropriate initial and boundary conditions must be supplied. In most wave problems, one must provide the initial conditions $\eta^*(x^*,0)$ and $u^*(x^*,0)$ (for tsunamis, it is usually assumed that $u^*(x^*,0) = 0$). The boundary condition far from the tsunami source area (``left boundary'') is
\begin{equation}\label{Run-up4}
\eta^*(x^*,t^*) \rightarrow 0, \quad u^*(x^*,t^*) \rightarrow 0 \quad (x^* \rightarrow - \infty).
\end{equation} 
If the tsunami source is far from the shore, it is convenient not to include the source area in the fluid domain and apply the following ``\emph{left}'' (incoming wave) boundary condition at some point $x^* = x_0^*$:
\begin{equation}\label{Run-up5}
u^*(x^*_0,t) = \sqrt{g/h(x^*_0)} \eta^*(x^*_0,t)   
\end{equation}
which corresponds to the tsunami wave approaching the shore. Another boundary condition is the boundedness of all functions on the unknown moving boundary,
\begin{equation}\label{Run-up6}
h(x^*) + \eta^*(x^*,t^*) = 0 \,,
\end{equation}
which determines the location of the moving shoreline. The condition (\ref{Run-up6}) is the main difference from the classical formulations of the Cauchy problem for hyperbolic systems.

There is an analytical method for solving this system, based on the use of the Riemann invariants. These invariants for a plane beach are
\begin{equation}\label{Run-up7}
  I_\pm = u^*\,\pm\,2\,\sqrt{\,g(- \alpha x^* + \eta^*)\,} + g\alpha t^*, 
\end{equation}
and the system (\ref{Riemann1}) can be rewritten as
\begin{equation}\label{Run-up8}
  \frac{\partial I_\pm}{\partial t^*}\,+\,\left(\,\frac{3}{4}\,I_\pm\,+
\,\frac{1}{4}\,I_\mp\,-\,g\alpha t^*\,\right)\,\frac{\partial I_\pm}{\partial x^*}\,=\,0.                  
\end{equation}
It is important to mention that this approach is applied for water waves on a beach of constant slope, and there are no rigorous results for arbitrary depth profiles $h(x^*)$. The existence of the Riemann invariants in the general case is an open mathematical problem.

Then the hodograph transformation can be applied to the system (\ref{Run-up8}), assuming that the determinant of the Jacobian $J\, = \,\partial\,(x^*,t^*)\,/\,\partial\,(I_+,I_-)$ does not vanish (this determinant vanishes when the wave breaks; we note that \textsc{Synolakis} (1987) \cite{Synolakis1987} has argued that this point is simply where the hodograph transformation becomes singular and the interpretation is that the wave breaks, and in fact corresponds to breaking during the rundown, at least for solitary waves). As a result, the following set of equations is derived:
\begin{equation}\label{Run-up9}
\frac{\partial x^*}{\partial I_\mp}\,-\,\left(\frac{3}{4}\,I_\pm\,+\,
\frac{1}{4}\,I_\mp\,-\,g\alpha t^*\,\right)\,\frac{\partial t^*}{\partial I_\mp}=0.
\end{equation}
These equations are still nonlinear but they can be reduced to a linear equation by eliminating $\,x^*\,(I_+,\,I_-)\,$:
\begin{equation}\label{Run-up10}
\frac{\partial^2 t^*}{\partial I_+\,\partial I_-}\,+\,\frac{3}{2\,(I_+\,-\,I_-)}
\,\left(\,\frac{\partial t^*}{\partial I_-}\,-\,\frac{\partial t^*}{\partial I_+}
\,\right)=0.                                   
\end{equation}
It is convenient to introduce the new variables
\begin{equation}\label{Run-up11}
\lambda\, = \,\frac{1}{2}\,(I_+\,+\,I_-)\,=\,u^*\,+\,g\alpha t^*,  \quad 
\sigma\, = \,\frac{1}{2}\,(I_+\,-\,I_-)\,=\,2\,\sqrt{\,g(- \alpha x^* + \eta^*)\,}.
\end{equation}
Then the system \eqref{Run-up10} takes the form
\begin{equation}\label{Run-up12}
\sigma\,\left(\,\frac{\partial^2 t^*}{\partial\lambda^2}\,-\,\frac{\partial^2 t^*}{\partial\sigma^2}\,\right)\,-\,3\,\frac{\partial t^*}{\partial\sigma}=0.
\end{equation}        
Expressing the time $t^*$ from Eq. \eqref{Run-up11},
\begin{equation}\label{Run-up13}
t^*\, = \,\frac{\lambda\,-\,u^*}{g\alpha},
\end{equation}
and substituting
\begin{equation}\label{Run-up14}
u^*\,=\,\frac{1}{\sigma}\,\frac{\partial\Phi}{\partial\sigma},
\end{equation}
we finally rewrite Eq. (\ref{Run-up12}) in the form of the classical cylindrical wave equation
\begin{equation}\label{Run-up15}
\frac{\partial^2\Phi}{\partial\lambda^2}\,-\,\frac{\partial^2\Phi}{\partial
\sigma^2}\,-\,\frac{1}{\sigma}\,\frac{\partial\Phi}{\partial\sigma}\,=\,0.
\end{equation}
All physical variables can be expressed through the function $\Phi(\sigma, \lambda)$. In addition to the time $t^*$ \eqref{Run-up13} and the velocity $u^*$ \eqref{Run-up14}, the horizontal coordinate $x^*$ and the water displacement $\eta^*$ are given by
\begin{eqnarray}\label{Run-up16}
x^* & = & \frac{1}{2g\alpha}\,\left(\,\frac{\partial\Phi}{\partial\lambda}\,-
\,u^{*2}\,-\,\frac{\sigma^2}{2}\right), \\
\eta^* & = & \frac{1}{2g}\,\left(\,\frac{\partial\Phi}{\partial\lambda}\,-
\,u^{*2}\,\right).\label{Run-up17}                  
\end{eqnarray}
So, the initial set of nonlinear shallow  water equations has been reduced  to  the  linear wave equation \eqref{Run-up15}  and all physical variables can be found via $\,\Phi\,$ using  simple  operations. The main advantage of this form of the nonlinear shallow-water system is  that the moving shoreline corresponds to $\,\sigma\,=\,0\,$ (since the total depth $\,h(x^*) + \eta^*(x^*,t^*) = 0\,$) and therefore Eq.~\eqref{Run-up15} is solved in the half-space $\,0\,\le\,\sigma\,<\,\infty\,$ with a fixed boundary, unlike the  initial equations.  The linear cylindrical wave equation~\eqref{Run-up15} is well-known in mathematical physics, and its solution can be presented in various forms (Green's function, Hankel and Fourier transforms). Using its solution, the wave field in ``\emph{physical}'' variables can be found from algebraic manipulations. Detailed analyses of the wave transformation and run-up have been performed for various initial conditions, see for instance \cite{CWY, Kanoglu2006, Tadepalli1996}.

Meanwhile, the typical situation in tsunamis is that the wave approaches the shore from deep water where the wave can be considered as linear. In this case it is possible to find the function $\Phi$ without using the implicit formulas of the hodograph transformation. Let us consider the linear version of the shallow water system: 
\begin{equation}\label{Run-up18} 
\frac{\partial u^*}{\partial t^*} + g \frac{\partial\eta^*}{ \partial x^*} = 0, 
\quad
\frac{\partial\eta^*}{\partial t^*}+ \frac{\partial}{\partial x^*}\,(-\alpha x^* u^*) = 0,   
\end{equation}
and apply the linearized version of the hodograph transformation
\begin{equation}\label{Run-up19}
\eta^* = \frac{1}{2g} \frac{\partial \Phi_l}{\partial \lambda_l},  \quad u^* = \frac{1}{\sigma_l} \frac{\partial \Phi_l}{\partial \sigma_l}, \quad x^* = - \frac{\sigma_l^2}{4 g \alpha}, \quad t^* = \frac{\lambda_l}{g \alpha},
\end{equation}
where the subscript $l$ denotes quantities derived from linear theory. In this case the system \eqref{Run-up18} reduces naturally to the same linear cylindrical wave equation
\begin{equation}\label{Run-up20}
\frac{\partial^2\Phi_l}{\partial\lambda^2_l}\,-\,\frac{\partial^2\Phi_l}{\partial
\sigma^2_l}\,-\,\frac{1}{\sigma_l}\,\frac{\partial\Phi_l}{\partial\sigma_l}\,=\,0,
\end{equation}
which has the same form as in nonlinear theory \eqref{Run-up15}. If the initial conditions for the wave field are determined far from the shoreline, where the wave is linear, then the initial conditions for both equations \eqref{Run-up15} and (\ref{Run-up20}) are the same, and therefore, their solutions will be the same,
\begin{equation}\label{Run-up21}
\Phi(\sigma, \lambda) = \Phi_l (\sigma_l, \lambda_l),
\end{equation} 
after replacing the arguments. So the function $\Phi$ can be found from linear theory.

From the operational point of view, it is important to know the extreme run-up characteristics like run-up height, rundown amplitude, onshore and offshore velocity, and these characteristics can be calculated within the framework of linear theory. This surprising result, also noted by \cite{Synolakis1987}, can be explained as follows. Indeed, it follows from Eq.~\eqref{Run-up21} that extreme values of $\Phi$ and its derivatives are the same. But for a moving shoreline ($\sigma = 0$) in extreme points of run-up or rundown, the velocity is zero, and the expressions of the hodograph transformations \eqref{Run-up17} and \eqref{Run-up19} coincide. So, it is believed that the extreme characteristics of tsunami run-up which determine the flooding zone can be found from linear theory despite the real nonlinear character of the wave process in the nearshore area, and this is an important result for tsunami engineering.

Moreover, the nonlinear dynamics of the moving shoreline ($\sigma = 0$) can be easily derived using linear theory. It follows from \eqref{Run-up13} that the moving shoreline velocity is
\begin{equation}\label{Run-up22}
u^*(\lambda, \sigma = 0)= \lambda - g \alpha t^*,
\end{equation}  
or in equivalent form
\begin{equation}\label{Run-up23}
u^*(t^*)= u_l^*(t^* + u^*/ g \alpha),
\end{equation}  
where the function $u_l^*(\lambda)$ is found using the known function $\Phi$. Therefore it can be found from linear theory (it is the velocity at the point $x^* = 0$). Similarly, the water displacement should be found first from linear theory (at the point $x^* = 0$)
\begin{equation}\label{Run-up24}
z_l^*(t^*) = \eta_l^*(x^*=0,t^*) = \alpha \int u_l^*(t^*)dt^*,
\end{equation}
and then one can find the ``\emph{real}'' nonlinear vertical displacement of the moving shoreline,
\begin{equation}\label{Run-up25}
z^*(t^*) = \eta^*(\sigma = 0) = \alpha \int u^*(t^*)dt^* = z_l^*(t^* + u^*/ \alpha g) - u^{*2}(t^*)/2g.
\end{equation}
As can be seen from these formulas, the extreme values of functions in linear and nonlinear theories coincide as we pointed it out already. The manifestation of nonlinearity is in the shape of the water oscillations on shore due to the nonlinear transformation \eqref{Run-up23}.

As an example, let us consider the run-up of monochromatic waves on the beach. It is enough to consider first the linear problem in the framework of the cylindrical wave equation \eqref{Run-up20}. The elementary bounded solution of this equation can be found in terms of Bessel functions:
\begin{equation}\label{shore1}
\eta^*(x^*,t^*) = \eta_R J_0 \left(\sqrt{\frac{4 \omega^2 |x^*|}{g \alpha}}\right) \cos\omega t^*,
\end{equation} 
with $\eta_R$ an arbitrary constant. Using asymptotic expressions for the Bessel function $J_0$ and matching with the solution of the mild slope equation (see \cite{Madsen2008}) one finds that the wave field far from the shoreline consists of the linear superposition of two waves propagating in opposite directions and having the same amplitudes (a standing wave):
\begin{equation}\label{shore2}
\eta^*(x^*,t^*) = 2\eta_0  \left(\frac{L}{|x^*|}\right)^{1/4} \cos\left(2\omega\sqrt{\frac{|x^*|}{g \alpha}} + \varphi \right)\cos\omega t^*,
\end{equation}
where the incident wave amplitude has been fixed to $\eta_0$ at the coordinate $x^*= - L$. The coefficient of wave amplification in the run-up stage is found to be
\begin{equation}\label{shore4}
\frac{\eta_R}{\eta_0} = 2\left(\frac{\pi^2 \omega^2 L}{g \alpha}\right)^{1/4} = 2\pi \sqrt{\frac{2L}{\lambda_0}},
\end{equation}
where $\lambda_0 = 2\pi \sqrt{g \alpha L}/\omega$ is the wavelength of the incident wave.

In their monograph \cite{Billingham2001}, Billingham and King use a different approach to find $\eta_R$. They suppose that at $x^*=-L$ there is an incident wave $\eta^*(-L,t^*)=2\eta_0\cos\omega t^*$. Matching the solution (\ref{shore1}) with it at $x^*=-L$ yields
\begin{equation}\label{shoreKing}
\eta^*(x^*,t^*) = 2\eta_0 J_0 \left(\sqrt{\frac{4 \omega^2 |x^*|}{g \alpha}}\right) /  J_0 \left(\sqrt{\frac{4 \omega^2 L}{g \alpha}}\right) \cos\omega t^*.
\end{equation} 
There is the possibility of a resonance, which occurs when $2\omega\sqrt{L/g\alpha}$ is a zero of $J_0$ and the solution \eqref{shoreKing} is then unbounded. Another way to look at this resonance is to consider Figure 9(a) in \cite{Madsen2008}. The resonance occurs when ones tries to force the wave amplitude to a finite value at one of the nodes of the solution \eqref{shore1}.

\section{A more realistic example}

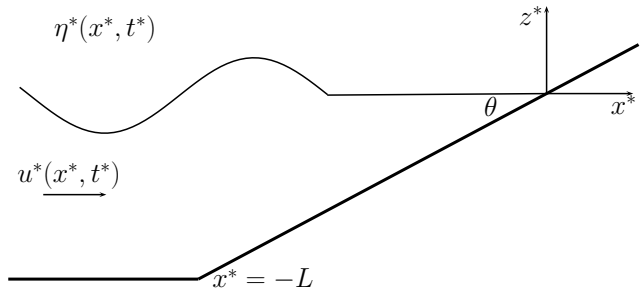
\begin{figure}
\begin{center}
\scalebox{0.5} 
{
\begin{pspicture}(-5.0,-3.64)(11.6,3.64)
\psline[linewidth=0.08cm](0.0,-3.62)(11.58,2.6)
\psline[linewidth=0.08cm](-5.0, -3.62)(0.0,-3.62)
\psline[linewidth=0.04cm](3.39,1.26)(9.18,1.3)
\infixtoRPN{+1.0*sin(0.8*(x+0.5)*180/3.1415)+1.25}
\psplot[linewidth=0.04]
     {-4.7}{3.4}{\RPN}
\psline[linewidth=0.04cm,arrowsize=0.05291667cm 2.0,arrowlength=1.4,arrowinset=0.4]{->}(9.16,1.3)(9.16,3.62)
\psline[linewidth=0.04cm,arrowsize=0.05291667cm 2.0,arrowlength=1.4,arrowinset=0.4]{->}(9.16,1.3)(11.44,1.3)
\usefont{T1}{ptm}{m}{n}
\rput(8.75,3.365){\LARGE $z^*$}
\usefont{T1}{ptm}{m}{n}
\rput(11.19,1.0){\LARGE $x^*$}
\usefont{T1}{ptm}{m}{n}
\rput(7.7,0.95){\LARGE $\theta$}
\psline[linewidth=0.04cm,arrowsize=0.05291667cm 2.0,arrowlength=1.4,arrowinset=0.4]{->}(-4.08,-1.38)(-2.42,-1.38)
\usefont{T1}{ptm}{m}{n}
\rput(-3.4,-0.85){\LARGE $u^*(x^*, t^*)$}
\usefont{T1}{ptm}{m}{n}
\rput(-2.45,3.0){\LARGE $\eta^*(x^*, t^*)$}
\rput(1.7,-3.62){\LARGE $x^* = -L$}
\end{pspicture} 
}
\end{center}
\caption{\em The geometry of a plane beach connected to a region of constant depth.}
\label{fig:matching}
\end{figure}

\begin{figure}
   \begin{center}
   \includegraphics[width=0.7\textwidth]{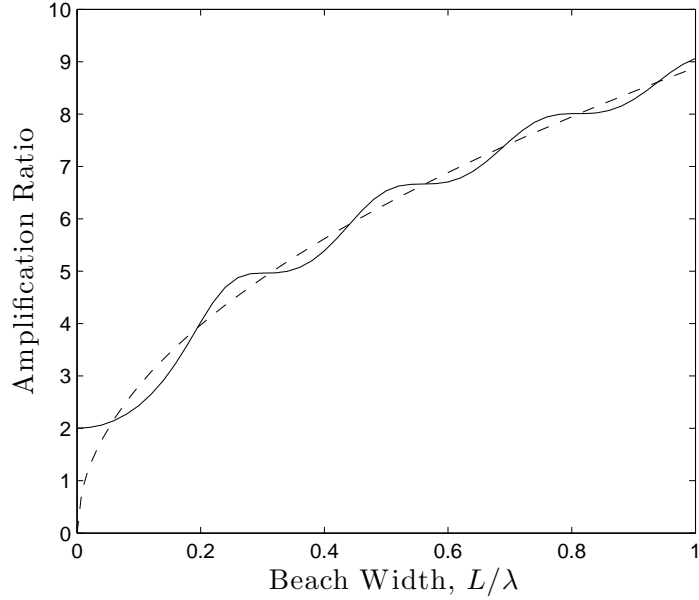}
   \end{center}
   \caption{\em Run-up height of a sine wave arriving from a basin of constant depth. The solid line is formula \eqref{shore38} and the dashed line is the result for an infinite beach of constant slope \eqref{shore4}.}
   \label{fig:keller}
\end{figure}

The rigorous theory described above is valid for the waves in a wedge of constant slope. For all other depth profiles rigorous analytical results are absent. Real bathymetries, which are complex in the ocean, can be approximated by a beach of constant slope in the vicinity of the shore only. If the ``\emph{matching}'' point is relatively far from the shoreline, the linear theory of shallow water can be applied for waves in a basin of complex bathymetry except in the nearshore area. Within this approximation, and arguing as in \cite{Synolakis1987}, the 1D linear wave equation
\begin{equation}\label{shore35}
\frac{\partial^2 \eta^*}{\partial t^{*2}} - \frac{\partial}{\partial x^*} \left( c^2(x^*) \frac{\partial \eta^*}{\partial x^*}\right) = 0,   \quad  c^2(x^*) = gh(x^*)
\end{equation}
should be solved analytically or numerically, and then its solution should be matched with the rigorous solution of the run-up problem described above. A popular example of such matching is given for the geometry presented in Fig.~\ref{fig:matching}, which is often realized in laboratory experiments. The elementary solution of the wave equation \eqref{shore35} for a basin of constant depth $h_0$ is the superposition of incident and reflected waves
\begin{equation}\label{shore37}
\eta^*(x^*,t^*) = \eta_0 \exp [\ui \omega (t^* - x^*/c)] + A_r \exp[\ui\omega (t^*+ x^*/c)]\ + \mbox{ c.c.}, \quad c = \sqrt{gh_0},
\end{equation}
with $\eta_0$ real and $A_r$ complex, and this solution should be matched with (\ref{shore1}) at the point $x^* = -L$ using the continuity of $\eta^*(x^*)$ and 
$d \eta^*/dx^*$. As a result, the unknown constants $A_r$ and $\eta_R$ can be calculated from the boundary conditions at $x^* = -L$, and the run-up amplitude is
\begin{equation}\label{shore38}
\frac{\eta_R}{\eta_0} = \frac{2}{\sqrt{J_0^2(\chi) + J_1^2(\chi)}},   \quad    \chi = \frac{2 \omega L}{c} = 4\pi\frac{L}{\lambda_0}.
\end{equation}
It is displayed in Fig.~\ref{fig:keller}. The solid line is formula \eqref{shore38} and the dashed line is the previous result for a beach of constant slope \eqref{shore4}. One can see that the agreement between both curves is quite good.

\section{Numerical Results}

The solutions described in the previous section are standing waves. If the motion starts from scratch, one does not have a standing wave at the beginning. A standing wave requires time to develop and during that time, runup amplification can be significant if the left boundary is a physical node in the developed standing wave solution. In their monograph \cite{Billingham2001}, \textsc{Billingham} and \textsc{King} suppose that at $x^* = -L$ there is an incident wave $\eta^*(-L,t^*) = 2\eta_0\cos\omega t^*$. Matching the solution \eqref{shore1} with it at $x^* = -L$ yields
\begin{equation}\label{shoreKing}
\eta^*(x^*,t^*) = 2\eta_0 J_0 \left(\sqrt{\frac{4 \omega^2 |x^*|}{g \alpha}}\right) /  J_0 \left(\sqrt{\frac{4 \omega^2 L}{g \alpha}}\right) \cos\omega t^*.
\end{equation} 
There is indeed the possibility of a resonance, which occurs when $2\omega\sqrt{L/g\alpha}$ is a zero of $J_0$ and the solution \eqref{shoreKing} is then unbounded. Another way to look at this resonance is to consider Figure 9(a) in \cite{Madsen2008}. The resonance occurs when ones tries to force the wave amplitude to a finite value at one of the nodes of the solution \eqref{shore1}. We will see below that this resonance can occur in numerical as well as laboratory experiments. 

\subsection{Waves on a plane beach}

\begin{figure}
  \subfigure[]{\includegraphics[width=0.49\textwidth]{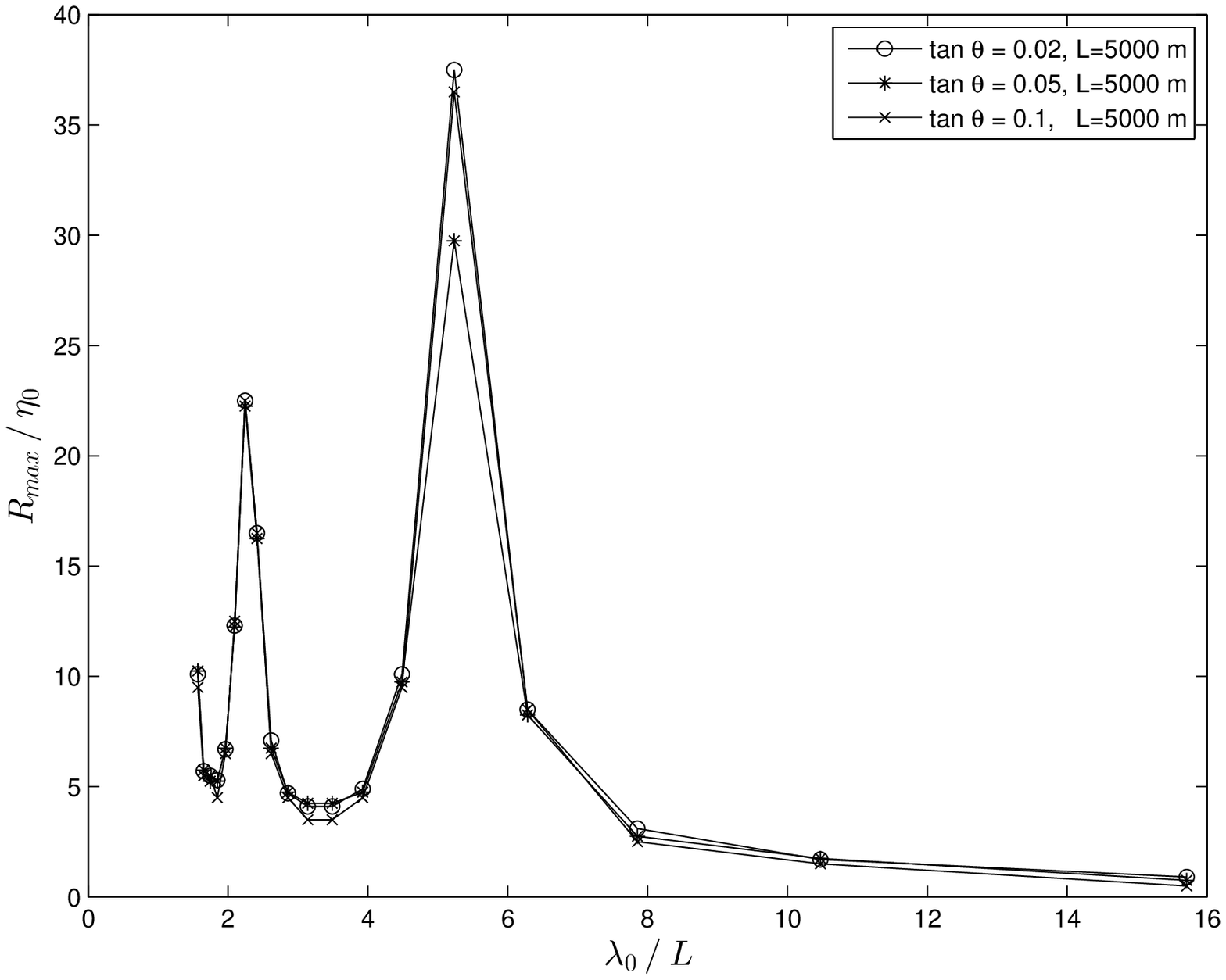}}
  \subfigure[]{\includegraphics[width=0.49\textwidth]{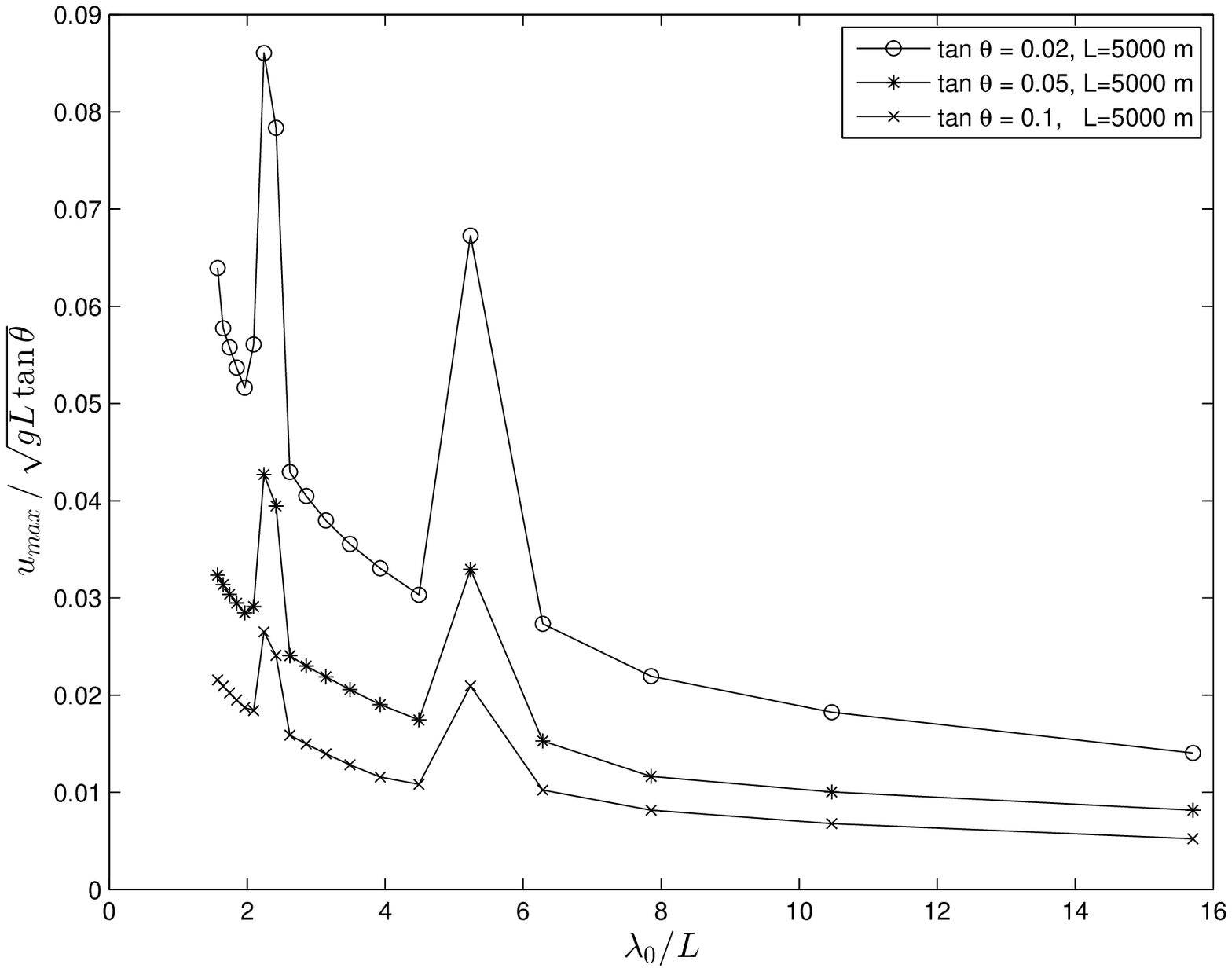}}
  \caption{\em Maximum run-up amplification $R_{\max}/\eta_0$ (a) and maximum horizontal velocity amplification (b) of monochromatic waves on a plane beach with respect to  nondimensional wavelength for three different slopes, namely $\tan{\theta} = 0.02\,;\, 0.05\,;\, 0.1$ ($L = 5000$ \m). Resonance is observed when the incoming wavelength is approximately $2.4$ and $5.2$ times the beach length.}
  \label{fig:PRL}
\end{figure}

\begin{figure}
   \begin{center}
   \includegraphics[width=0.79\textwidth]{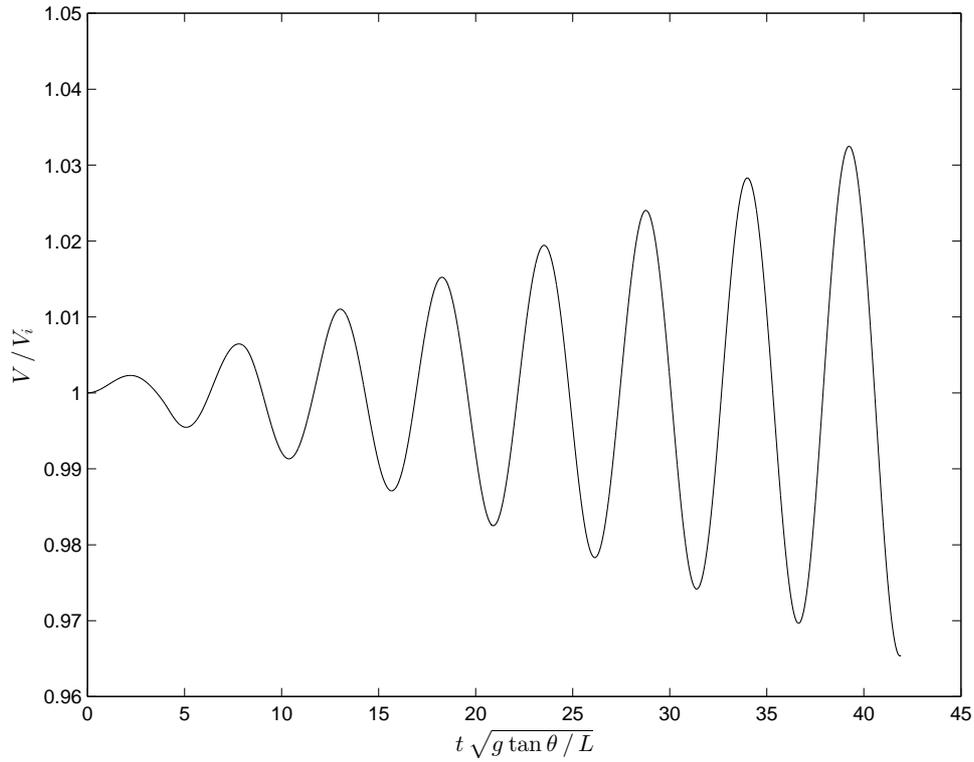}
   \end{center}
   \caption{\em Evolution of the volume of fluid $V$ inside the computational domain during resonance ($\alpha = 0.02$, $L = 5000$ \m, $\lambda_0/L =5.2$). $V_i$ is the initial volume.}
   \label{fig:volume}
\end{figure}

We present some additional results on the resonant long wave run-up phenomenon on a plane beach described by \textsc{Stefanakis} \emph{et al.} (2011) \cite{Stefanakis2011}. Namely, we look at the maximum run-up amplification of monochromatic waves, but this time we use milder slopes, which are more geophysically relevant ($\tan{\theta} = 0.02\,;\, 0.05\,;\, 0.1$). For our simulations we used the NSWE in one dimension, which were solved numerically by a Finite Volume Characteristic Flux scheme with UNO2 type of reconstruction for higher order terms and a third order Runge-Kutta time discretization. The left boundary condition is implemented as in \cite{Ghidaglia2005}. The model is described in detail and validated by \textsc{Dutykh} \emph{et al.} (2011) \cite{Dutykh2011e}. Monochromatic forcing ( $\eta^*(-L, t^*) = 2\eta_0 \cos\omega t^*$ ) on an infinite sloping beach was found to lead to resonant run-up by non-leading waves (see Fig.~\ref{fig:PRL}(a)) when the nondimensional  wavelength is $\lambda_0 / L \approx 5.2$, where $\lambda_0 = 2 \pi \sqrt{g \alpha L}/ \omega$ is the incident wavelength, which comes as a direct consequence of Eq.~\eqref{shoreKing} when the wave at the seaward boundary is specified, since resonant states can be identified with the roots of the Bessel function $J_0$. Resonance is also observed for the maximum horizontal velocities assumed by the waves (Fig.~\ref{fig:PRL}(b)), which are found at the shoreline. Since
\begin{equation*}
\eta^*(x^*,t^*) = \eta_R J_0(\sigma) \cos{\omega t^*}\ ,\ u^*(x^*,t^*) = -\frac{2 \omega\eta_R}{\sigma\tan{\theta}} J_1(\sigma) \sin{\omega t^*}
\end{equation*}
where
\begin{equation*}
\sigma = 2\omega\, \sqrt{\frac{|x^*|}{g \tan{\theta}}}
\end{equation*}
and
\begin{equation*}
\lim_{\sigma \rightarrow 0} \frac{J_1(\sigma)}{\sigma} = \frac{1}{2},
\end{equation*}
then the shoreline velocity is
\begin{equation}\label{shoreline_velocity}
u_s^*(t^*) = -\frac{\omega \eta_R}{\tan{\theta}} \sin{\omega t^*},
\end{equation}
which implies that when the run-up is resonant, so is the shoreline velocity. The resonance mechanism was found to rely on a synchronization between incident and receding waves but should be distinguished from wavemaker resonance since the computational domain is not closed (Fig.~\ref{fig:volume}), as it would be in a laboratory setting, and we can observe strong inflow-outflow during run-up and run-down. Furthermore, the experiments of \cite{Ezersky2012} confirmed this claim, by observing that the resonant frequency of the system is different from the frequency that leads to the resonant run-up.

\begin{figure}
  \centering
  \subfigure[]{\includegraphics[width=0.49\textwidth]{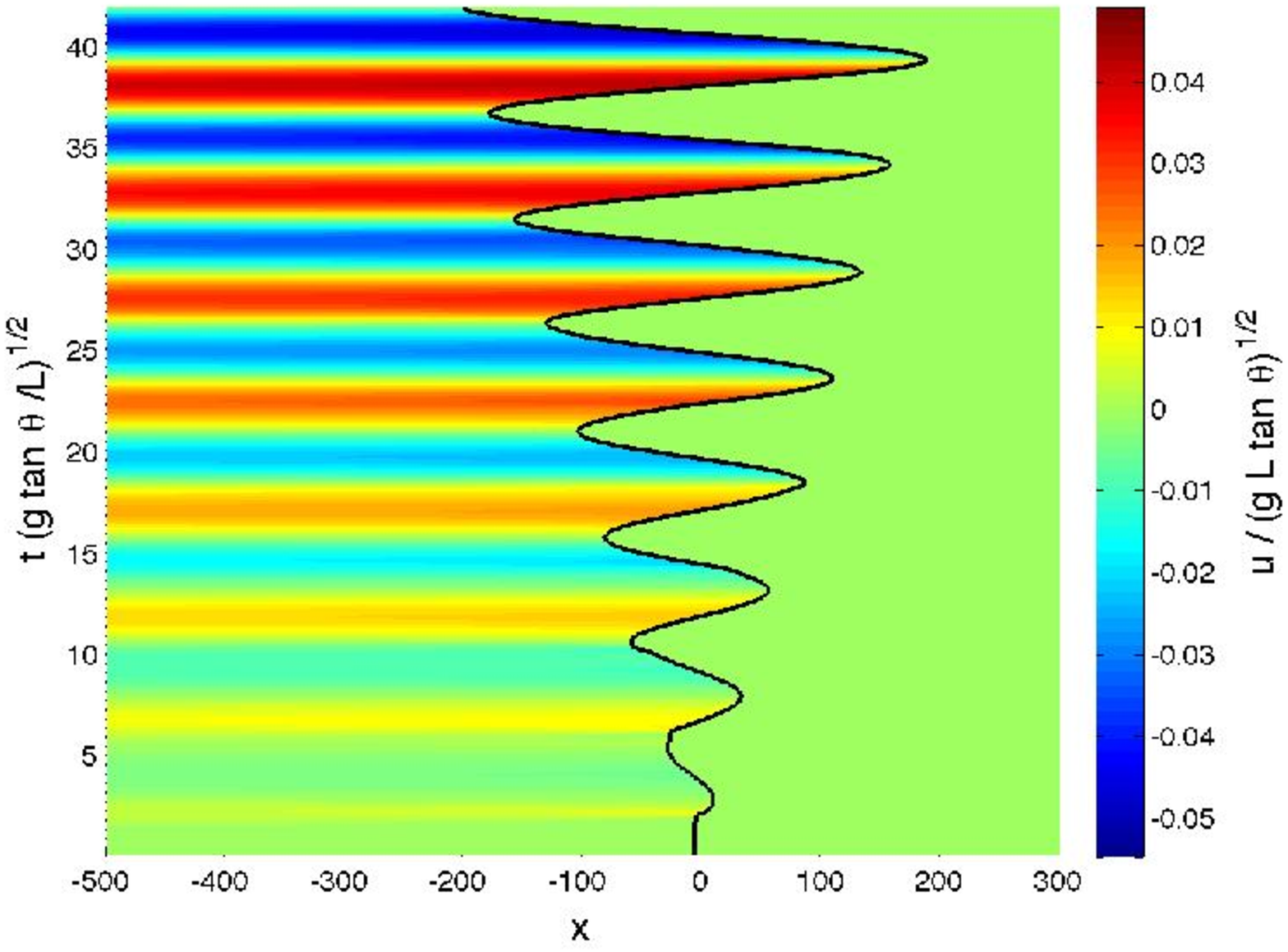}}
  \subfigure[]{\includegraphics[width=0.49\textwidth]{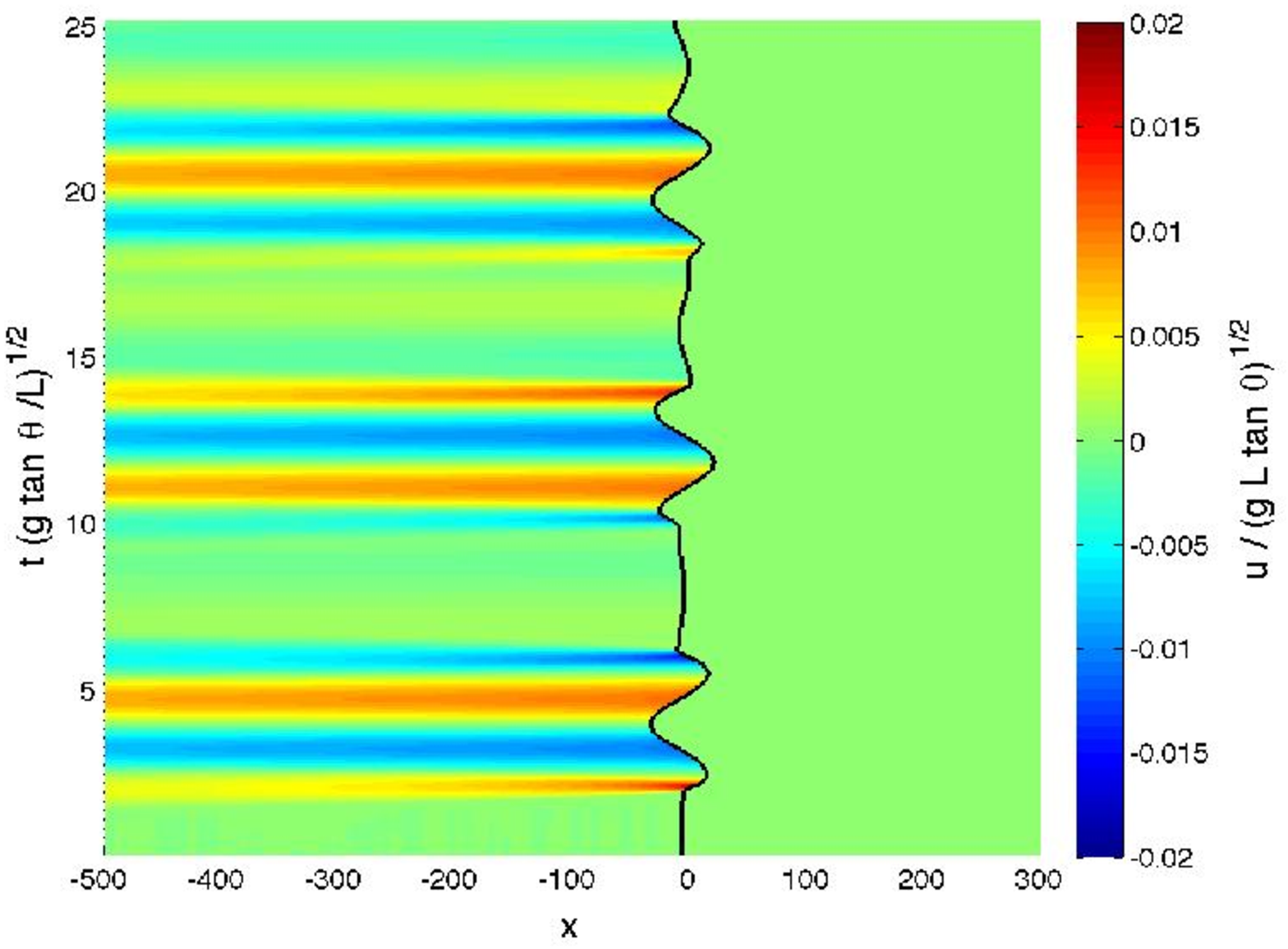}}
  \caption{\em Spatio-temporal behaviour of non-dimensional horizontal velocity $u/(g \tan{\theta}L)^{1/2}$ in the resonant regime (a) and non-resonant regime (b). The black line describes the evolution of the shoreline position in time. In both cases $\tan{\theta} = 0.05$ and $L = 5000$ \m.}
\label{fig:SpatVel}
\end{figure}

The spatio-temporal behavior of the non-dimensional horizontal velocity is shown in Fig.~\ref{fig:SpatVel}. For visualization purposes, we plot only the last $500$ \m of the beach to the left of the initial shoreline position. We can observe that in the resonant regime, after the rundown induced by the leading wave, during run-up of subsequent waves, a fixed spatial point undergoes an abrupt change of velocity, from highly negative to highly positive values. Furthermore, the maximum absolute velocity increases over time in the resonant regime, while this is not true in the non-resonant case. To give a feeling of dimensions, imagine a plane beach with $\tan{\theta} = 0.01$ and an incoming wave of amplitude $\eta_0 = 1$ \m at $L = 10,000$ m offshore where the water depth is $h_0 = 100$ \m. For that wave to be in the resonant regime, its wavelength has to be approximately $52,000$ \m. If there is a run-up amplification close to $40$, this means that $\max{u_s} \approx 15$ \nicefrac{\m}{\s} according to Eq.~\eqref{shoreline_velocity}.

\begin{figure}
  \centering
  \includegraphics[width=0.7\textwidth]{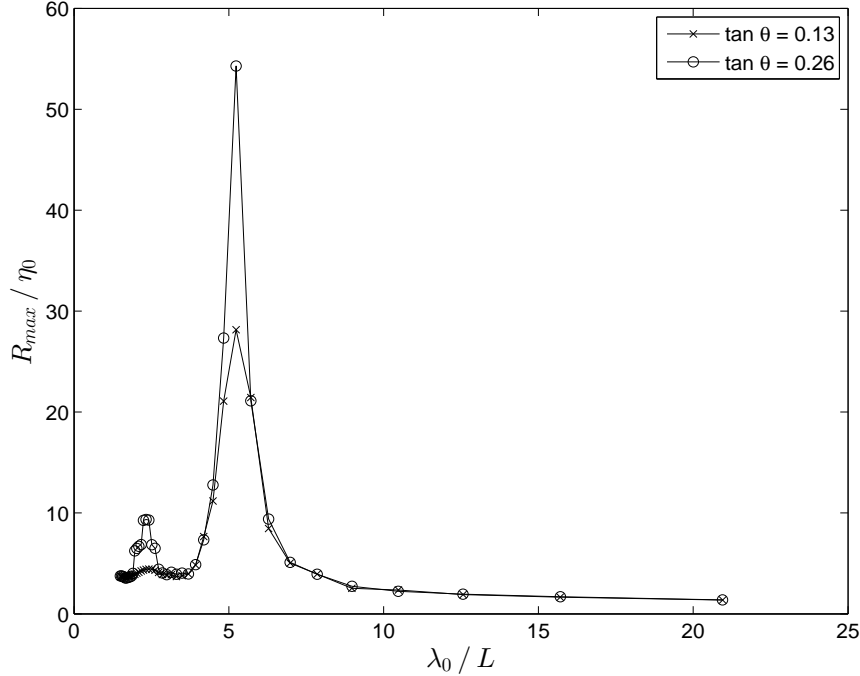}
  \caption{\em Maximum run-up of monochromatic waves on a plane beach as a function of nondimensional wavelength for two different slopes, namely $\tan\theta = 0.13 ; 0.26$ ($L = 12.5$ \m). The results were obtained with VOLNA code, a 2D finite volume solver of the NSWE.}
   \label{fig:VOLNA}
\end{figure}

In order to increase our confidence in the numerical solver that we use, we ran simulations using the VOLNA code \cite{Dutykh2009a}, which is a NSWE solver in two horizontal dimensions. VOLNA has been validated with the Catalina benchmark problems \cite{noaa_report}, which are established by the National Oceanic and Atmospheric Administration (NOAA) Center for Tsunami Research and consist of a series of test cases based on analytical, experimental and field observations. We tested the maximum run-up on plane beaches with slopes $\tan\theta = 0.13; 0.26$ when the beach length is $L = 12.5$ \m. We used a smaller beach length in order to limit the computational cost of the 2D simulations and we chose to follow the same setup as \cite{Stefanakis2011}. The results (Fig.~\ref{fig:VOLNA}) are in good agreement with the results obtained before (Fig.~2 in \cite{Stefanakis2011}) and the same resonant regime is observed again. Consequently, due to the confidence in the results we obtained and the reduced computational cost, we decided to continue working with the NSWE solver in one horizontal dimension.

\begin{figure}
   \begin{center}
   \includegraphics[width=0.89\textwidth]{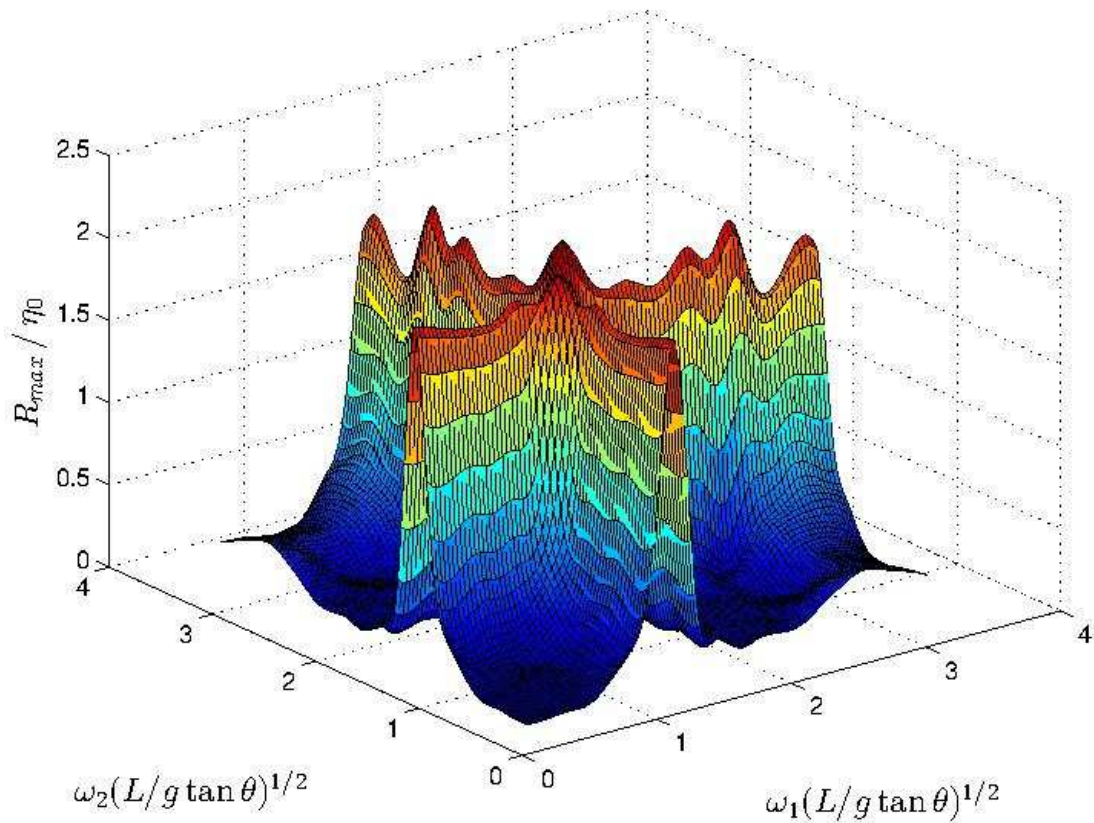}
   \end{center}
   \caption{\em Maximum run-up of bichromatic waves with respect to nondimensional frequency ($\tan\theta = 0.13\ ,\ L = 12.5$ \m).}
   \label{fig:bichromatic}
\end{figure}

In order to further investigate the effects of modal interactions in the resonant regime, we tested incoming waves of bichromatic modal structure. In order to be consistent with the monochromatic case, each mode had half the amplitude of the equivalent monochromatic wave ($\eta_0$). Our computations (Fig.~\ref{fig:bichromatic}) show that no important new interactions occur. When one of the two frequencies is resonant, the run-up is dominated by it, while the other does not alter the dynamics. If both frequencies are resonant, their constructive interference is small overall and does not differ significantly from the equivalent monochromatic resonant state. Therefore, this result indicates that the resonant run-up mechanism is robust and is not restricted to monochromatic waves only.

\begin{figure}
   \begin{center}
   \includegraphics[width=0.79\textwidth]{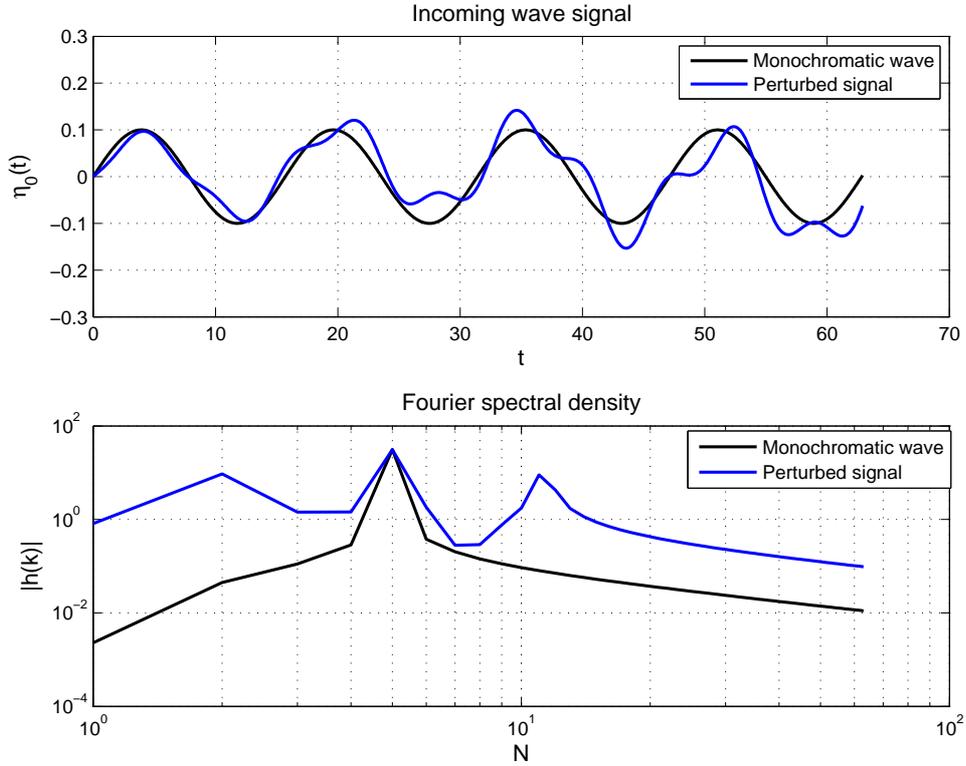}
   \end{center}
   \caption{\em Comparison of a typical monochromatic wave and a corresponding semi-randomly perturbed signal both in physical space (top) and Fourier space (bottom), where $N$ is the Fourier mode and $h(k)$ is the spectral amplitude. Time $t$ is in seconds and the free-surface elevation $\eta_0(t)$ is in meters.}
   \label{fig:random}
\end{figure}

\begin{figure}
   \begin{center}
   \includegraphics[width=0.79\textwidth]{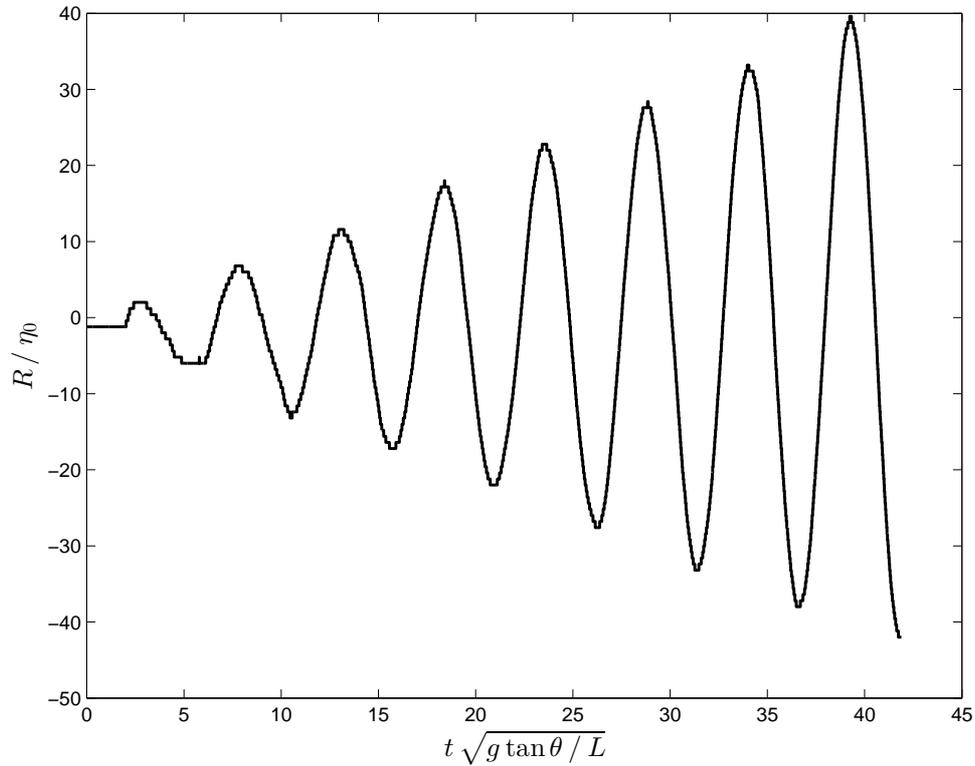}
   \end{center}
   \caption{\em Run-up timeseries of a perturbed resonant monochromatic wave when the slope is $0.02$ and $L = 5000$ \m.}
   \label{fig:randomRun-up}
\end{figure}

In order to further investigate the robustness of the resonant run-up mechanism, we introduced $10$ semi-random perturbation components to the monochromatic wave signal.  The amplitude of the perturbations followed a normal distribution with zero mean and standard deviation much less than the wave amplitude $\eta_0$. The perturbation frequency followed the lognormal distribution. The monochromatic wave and a corresponding semi-randomly perturbed signal in physical and Fourier spaces are shown in Fig.~\ref{fig:random}. By running the simulation in the resonant regime when the slope $\alpha = 0.02$ and $L = 5000$ \m we obtained the same run-up timeseries that we would obtain with the unperturbed monochromatic wave (Fig.~\ref{fig:randomRun-up}). Therefore we can conclude that the resonant run-up mechanism is robust and the resonant frequency dominates the run-up.

\subsection{Piecewise linear bathymetry}

\begin{figure}
   \begin{center}
   \includegraphics[trim=0.0cm 2.3cm 0.0cm 0.0cm, clip=true, width=0.79\textwidth]{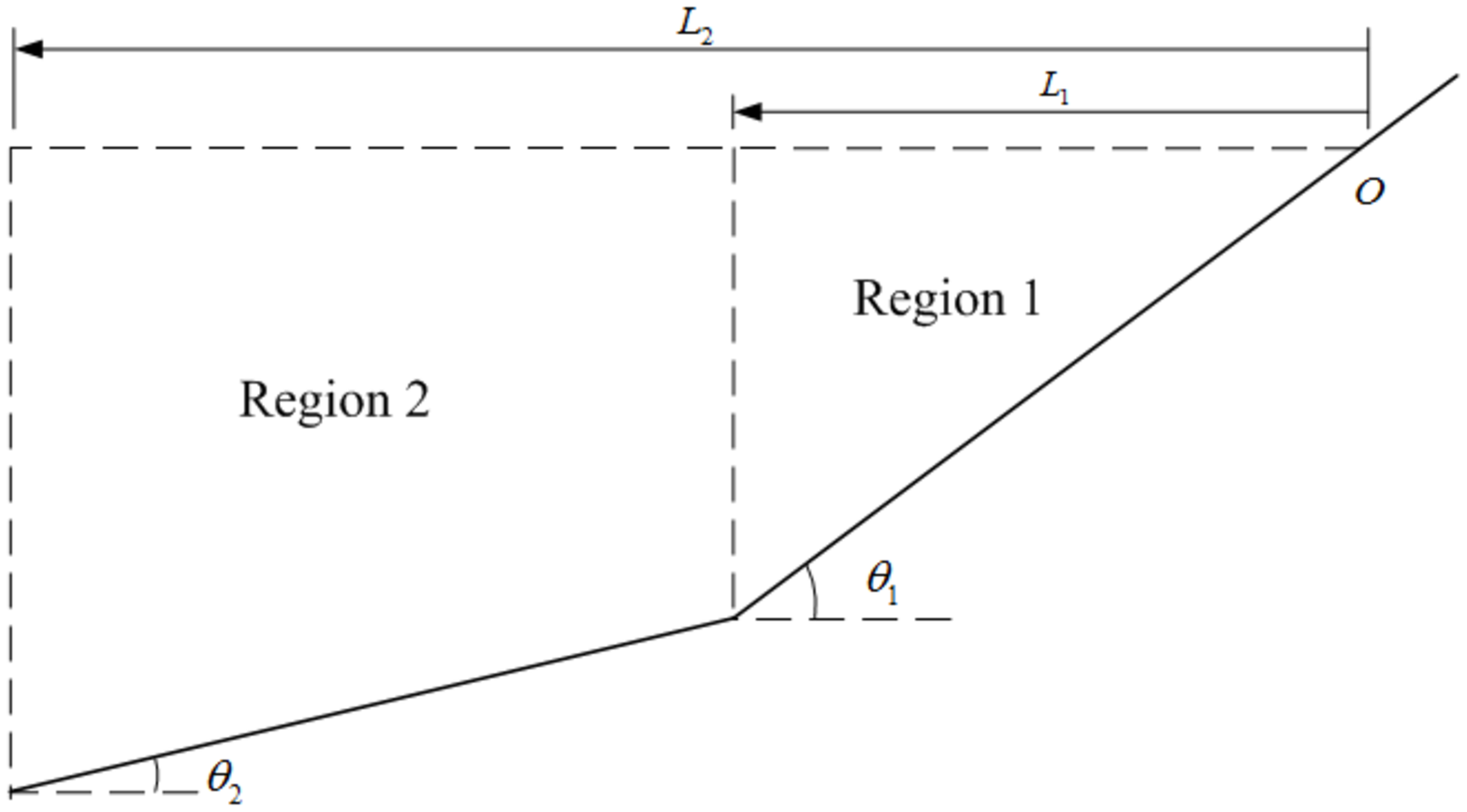}
   \end{center}
   \caption{\em Schematic of the piecewise linear bathymetry.}
   \label{fig:2slopes}
\end{figure}

\textsc{Kanoglu} \& \textsc{Synolakis} (1998) \cite{Kanoglu1998} developed a general methodology to study the problem of long wave run-up over a piecewise linear bathymetry and applied it to the study of solitary wave run-up, but in their formulation, the last offshore segment of the bathymetry consisted of a flat bottom. Here we will only use two uniformly sloping regions as in Fig.~\ref{fig:2slopes}. Following the steps of \textsc{Lamb} (1932) \cite{Lamb1932}, we take the linearized form of the NSWE \eqref{Riemann1} and search for solutions of the type
\begin{equation}\label{Lamb}
\eta^*(x^*,t^*) = Z(x^*) \cos{\omega t^*}, \quad 
u^*(x^*,t^*) = V(x^*) \sin{\omega t^*}.
\end{equation}
By inserting Eq.~\eqref{Lamb} into Eq.~\eqref{Riemann1} we obtain
\begin{equation}\label{BesselEq}
h(x^*)\frac{\d^2 Z}{\d x^{*2}} + \frac{\d h(x^*)}{\d x^*} \frac{\d Z}{\d x^*} + \frac{\omega^2 Z}{g} = 0,
\end{equation}
\begin{equation}\label{VelEq}
  V^* = -\frac{g}{\omega} \frac{\d Z}{\d x^*}.
\end{equation}
Since the bathymetry is piecewise linear, at each segment we have $h_i(x^*) = -\alpha_i x^* + c_i$, where $\alpha_i = \tan{\theta_i} \neq 0$, $c_i$ is a constant and the subscript $i$ is indicative of the segment number from the shoreline to the seaward boundary. In that case, Eq.~\eqref{BesselEq} becomes a standard Bessel equation of order zero and the general solution for $Z_i$ and $V_i$ is
\begin{equation}\label{BesselSol}
  Z_i(x^*) = A_i J_0(\sigma) + B_i Y_0(\sigma), \quad V_i(x^*) =  \frac{2 \omega}{\sigma \alpha_i} \frac{\d Z_i}{\d \sigma},
\end{equation}
where $A_i$ and $B_i$ are linear coefficients, $J_n$ and $Y_n$ are the $n$th order Bessel functions of the first and second kind respectively and
\begin{equation*}
  \sigma = \frac{2\omega}{\sqrt{g}} \sqrt{\frac{-x^*+\frac{c_i}{\alpha_i}}{\alpha_i}}.
\end{equation*}

\begin{figure}
   \begin{center}
   \includegraphics[width=0.59\textwidth]{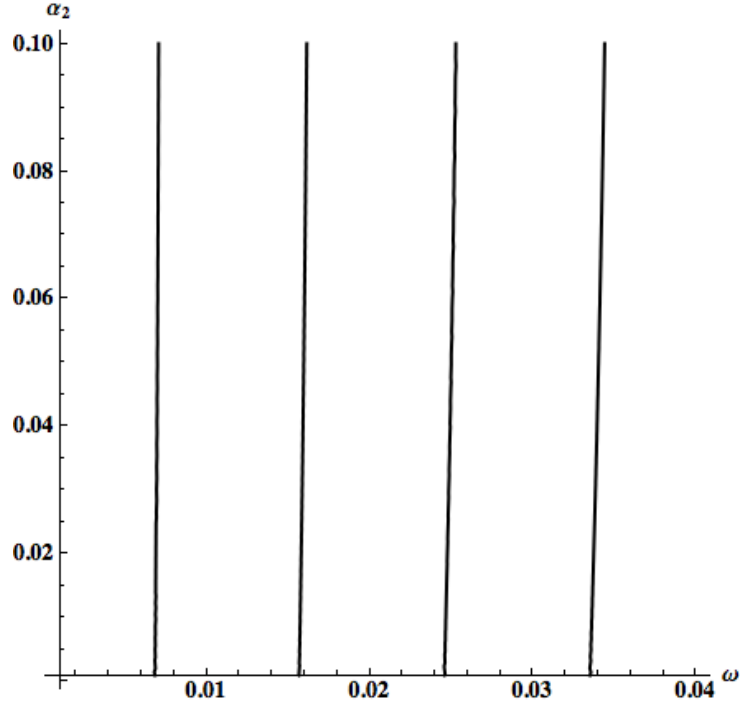}
   \end{center}
   \caption{\em Plot of the zeros of the determinant in the denominator of Eq.~\eqref{Runup2Slopes} as a function of $\omega$ and $\alpha_2$ when $\alpha_1 = 0.02,\ L_1 = 5000$ \m and $L_2 = 6000$ \m.}
   \label{fig:ZeroNom}
\end{figure}

\begin{figure}
   \begin{center}
   \includegraphics[width=0.79\textwidth]{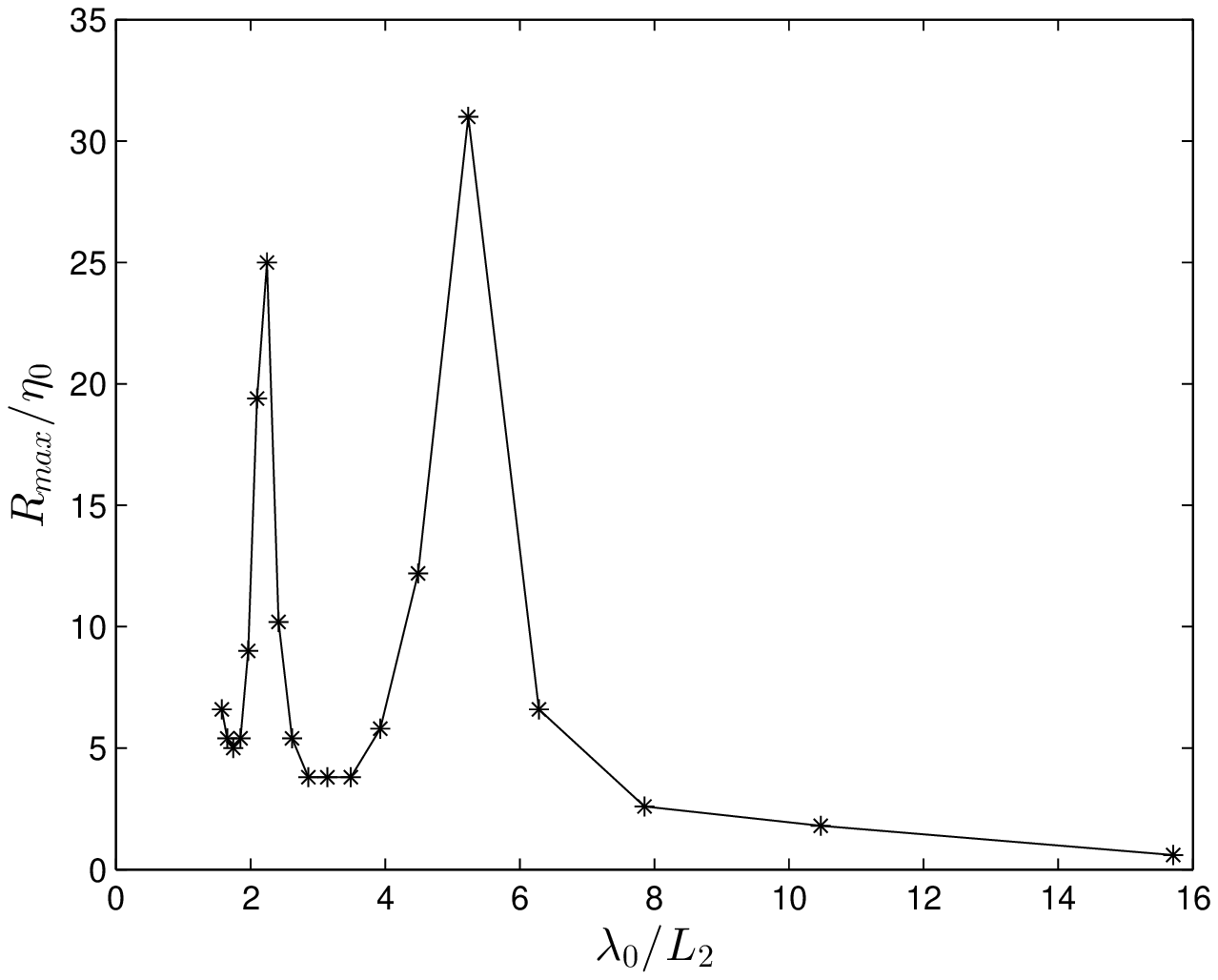}
   \end{center}
   \caption{\em Run-up amplification of monochromatic waves on a piecewise linear bathymetry consisting of two segments as a function of the nondimensional wavelength when $\alpha_1 = \tan{\theta_1} = 0.02, \alpha_2 = \tan{\theta_2} = 0.01, L_1 = 5000$ \m and $L_2 = 6000$ \m.}
   \label{fig:2slopeResonance}
\end{figure}

In order to solve this problem we require continuity of the free surface elevation and of the horizontal fluxes at two adjacent segments and we prescribe a wave amplitude $Z_i= \eta_0\cos\omega t^*$ at the seaward boundary. Here for simplicity we will focus on the case of two segments but it can be generalized to an arbitrary number of segments as it is shown in \cite{Kanoglu1998}. In the first segment boundedness of the free surface elevation at the shoreline ($x^* = 0$) requires $B_1 = 0$. Since $J_0(0) = 1$, $A_1$ represents the run-up and therefore we will name it $\eta_R$. Hence, we have the following linear system of equations:
\begin{eqnarray}
J_0(\sigma_1) \eta_R - J_0(\sigma_2) A_2 - Y_0(\sigma_2) B_2 &=& 0 \\
J_1(\sigma_1) \eta_R - J_1(\sigma_2) A_2 - Y_1(\sigma_2) B_2 &=& 0 \\
J_0(\sigma_3) A_2 + Y_0(\sigma_3) B_2 &=& \eta_0,
\end{eqnarray}
where
\begin{eqnarray}
\sigma_1 &=& 2 \omega \sqrt{\frac{L_1}{g \alpha_1}} \\
\sigma_2 &=& \frac{2 \omega}{\alpha_2} \sqrt{\frac{\alpha_1 L_1}{g}} \\
\sigma_3 &=& \frac{2 \omega}{\alpha_2} \sqrt{\frac{\alpha_2 (L_2 - L_1) + \alpha_1 L_1}{g}}.
\end{eqnarray}
Then
\begin{equation}\label{Runup2Slopes}
\eta_R = \left| \begin{array}{ccc} 0 & - J_0(\sigma_2) & - Y_0(\sigma_2) \\ 0 & - J_1(\sigma_2) & - Y_1(\sigma_2) \\ \eta_0 & J_0(\sigma_3) & Y_0(\sigma_3) \\ \end{array} \right| / \left| \begin{array}{ccc} J_0(\sigma_1) & - J_0(\sigma_2) & - Y_0(\sigma_2) \\ J_1(\sigma_1) & - J_1(\sigma_2) & - Y_1(\sigma_2) \\ 0 & J_0(\sigma_3) & Y_0(\sigma_3) \\ \end{array} \right|.
\end{equation}
Therefore, when the determinant in the denominator vanishes, the run-up becomes resonant (Fig.~\ref{fig:ZeroNom}). Furthermore, the shoreline velocity is now given by Eq.~\eqref{BesselSol} as
\begin{equation}\label{ShoreVel}
  V_s = \lim_{\sigma \to 0}\, -\frac{2 \omega \eta_R}{\sigma \alpha_1}\, J_1(\sigma) \ \Rightarrow V_s = -\frac{\omega}{\alpha_1} \eta_R,
\end{equation}
which indicates that when the run-up is resonant, so is the shoreline velocity.\footnote{During run-up,  the maximum shoreline velocity is not reached when the wave reaches its maximum run-up. So the joint resonance is not as obvious as it may look. In the literature, there are much less results on velocities than on run-up. One exception is the paper by \cite{Madsen2008}.} The same argument of course applies to the case of the infinite sloping beach (see Fig.~\ref{fig:PRL}b). Numerical simulations performed in this setting with $\alpha_1 = 0.02$, $\alpha_2 = 0.01$, $L_1 = 5000$ \m and $L_2 = 6000$ \m ($\eta_0 = 0.1$ \m) agree with the above analytical solution and again resonant wavelengths can be identified (Fig.~\ref{fig:2slopeResonance}).

\subsection{Plane beach connected to a flat bottom}

A more characteristic bathymetric profile consists of a constant depth region connected to a sloping beach, hereafter referred as the canonical case (Fig.~\ref{fig:matching}). Using this profile \textsc{Madsen} \& \textsc{Fuhrman} (2008) \cite{Madsen2008} showed very good agreement between theory and their computations for a range of wavelengths  $1 < \lambda_0 / L < 7$ and wave nonlinearity $0.001 < \eta_0 / h_0 < 0.01$, in which even a tsunami at $h_0 = 100$ \m is fairly linear.\footnote{With $h_0 = 100$ \m, the corresponding interval for $\eta_0$ is  $0.1$ \m $< \eta_0 < 1$ \m.}  For their computations, they placed a relaxation zone close to the wave generation area. This relaxation zone is basically a filtering over one wavelength of the computed free surface elevation and the free surface elevation induced only by the incoming wave, which is an idea borrowed from the widely used sponge layers, but at the same time it is significantly different due to its placing between the wave generation area and the region of interest. It is applied so that no reflected waves from the beach interact with the forcing boundary because there is no clear understanding of how to impose both incoming and outgoing waves at a boundary. It is convenient because it allows for a reduction of the computational cost and has been used successfully in several other studies (e.g. \cite{Mayer1998, Madsen2002, Lu2007}) but it is somewhat artificial. The length of the relaxation zone should be comparable to the wavelength, but the resonant wavelength is found to be greater than the beach length, which is the reason why we could not employ it in the infinite slope case.

\begin{figure}
   \begin{center}
   \includegraphics[width=0.75\textwidth]{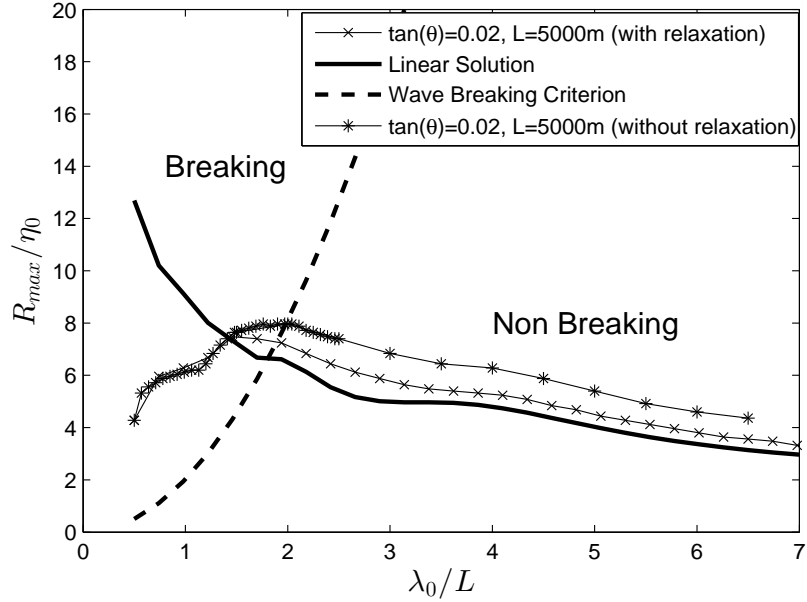}
   \end{center}
   \caption{\em Maximum run-up amplification as a function of nondimensional wavelength for the canonical case ($\eta_0 = 1.25$ \m , $h_0 = 100$ \m, $\tan \theta = 0.02$).}
   \label{fig:CanonicalRun-up_002}
\end{figure}

For the current bathymetric profile, our objectives were to investigate both if resonance would occur and whether the existence of the relaxation zone would play any role on the run-up. Hence we examined the run-up of monochromatic waves of amplitude $\eta_0 = 1.25$ \m on a plane beach with slope $\tan\theta = 0.02$, which reached a maximum depth $h_0 = 100$ \m. When we inserted a relaxation zone, one wavelength long, the constant depth region was two wavelengths long. In the other case where we avoided the use of relaxation we wanted to ensure that no reflected waves would reach the left boundary during the simulations. To achieve that, we sent four waves and the constant depth region was four wavelengths long, while the final time was set to eight wave periods. This type of setup is more natural since no artificial filtering is used but at the same time is more computationally demanding, due to the double length of the constant depth region. In Fig.~\ref{fig:CanonicalRun-up_002} we observe that both with and without the relaxation zone, the computations predict slightly higher maximum run-up values than the ones predicted by the theory in the non-breaking regime, but the qualitative behavior is the same.

The discrepancies between theory and computations are higher when the use of a relaxation zone is avoided. Like \textsc{Pelinovsky} (1992) \cite{Pelinovsky1992}, one can introduce the breaking number $Br=\omega^2\eta_R/g\alpha^2$. When $Br = 1$, or $\eta_R / \eta_0 = g \tan^2 \theta / (\eta_0 \, \omega^2)$,  the analytical solution breaks down. When $Br > 1$, the wave breaks. According to \cite{Mei2005}, this criterion can only be used as a qualitative criterion. When waves are close to breaking, the run-up amplification reaches its maximum. However, we cannot observe any significant resonance as we did in the infinite sloping beach example \cite{Stefanakis2011}. Wave breaking in the context of NSWE is demonstrated by the creation of a very steep wavefront and actually it is a common practice in tsunami modeling, for people who use Boussinesq systems, to switch to NSWE as soon as the slope of the wavefront exceeds a threshold (e.g. \cite{shi2012, Tissier2012}). Above the breaking threshold, we observe in Fig.~\ref{fig:CanonicalRun-up_002} that theory and computations do not agree. However, the computations qualitatively follow the trend of the laboratory experiments presented by \textsc{Ahrens} (1981) \cite{Ahrens1981} even though they refer to irregular wave run-up.

It is well known that in the context of NSWE, as waves propagate over a flat bottom, the wavefront tends to become steeper and the higher the wave nonlinearity, the faster the wave steepening. \textsc{Synolakis} \& \textsc{Skjelbreia} (1993) \cite{Synolakis1993} have showed that while, offshore and far from breaking the wave evolves with Green's law, closer to breaking the evolution is more rapid, and they named it the Boussinesq regime. In the previous case the discrepancies observed when using the relaxation zone and when it is not used, could be attributed to the different lengths of the constant bottom region (hereafter $L_0$). Before, the wave nonlinearity was $\eta_0 / h = 0.0125$ and in order to increase the effect of wave steepening we decided to double the incoming wave amplitude. Hence, we tested three different cases, namely the same two as before, one without a relaxation zone and $L_0 = 4\lambda_0$, one with relaxed boundary condition and $L_0 = 2\lambda_0$ and finally one with a relaxation zone but now the constant depth region has a length equal to 4 wavelengths.

\begin{figure}
   \begin{center}
   \includegraphics[width=0.99\textwidth]{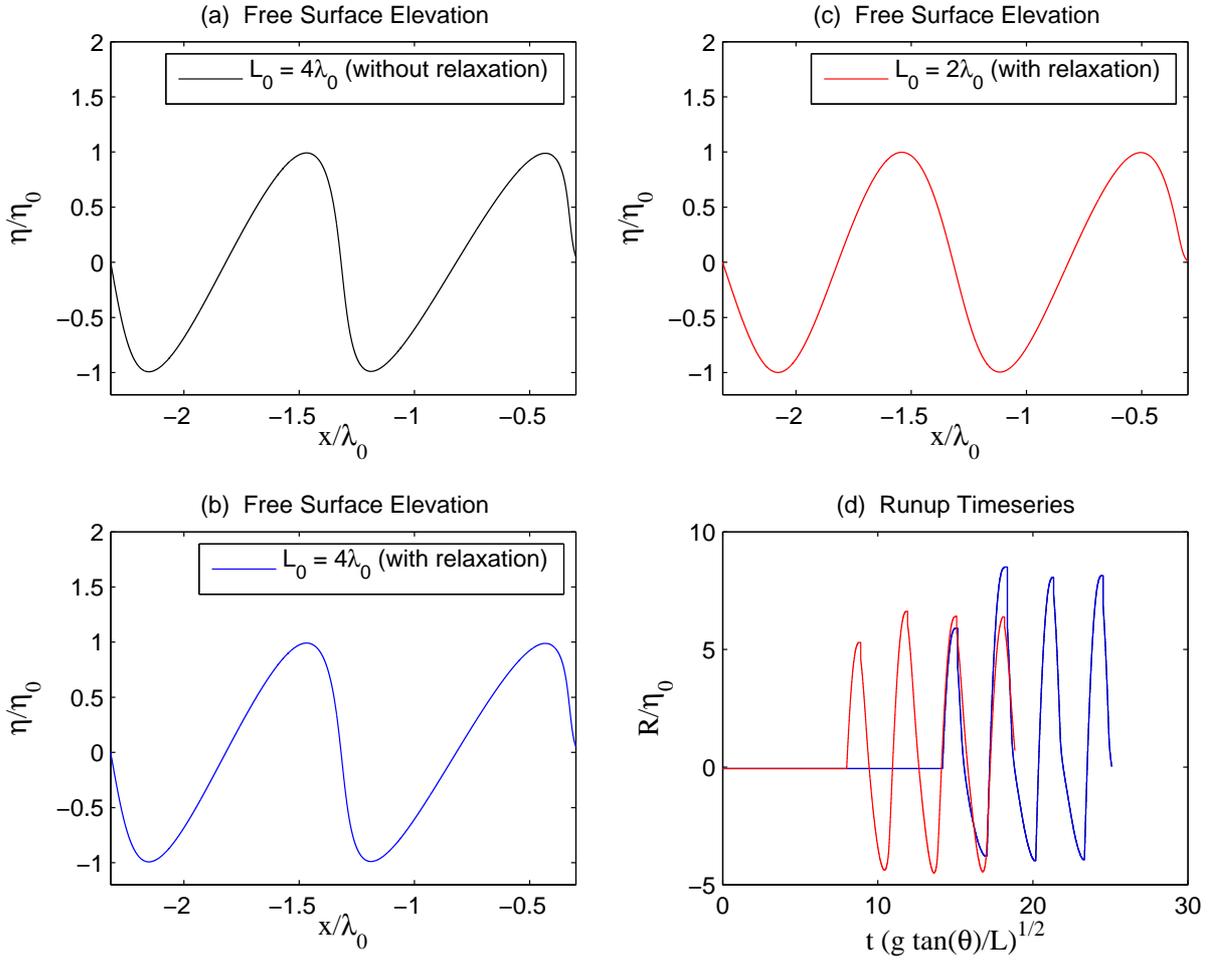}
   \end{center}
   \caption{\em Snapshots of free surface elevation over the constant depth region (a,b,c). The horizontal extent is two wavelengths offshore from the toe of the beach ($\lambda_0 / L = 3.14$, $\eta_0 =2.5$ \m). Steeper wavefronts are observed when $L_0 = 4\lambda_0$. Run-up timeseries (d). Waves with steeper wavefront run-up higher.}
   \label{fig:Steep}
\end{figure}

\begin{figure}
   \begin{center}
   \includegraphics[width=0.7\textwidth]{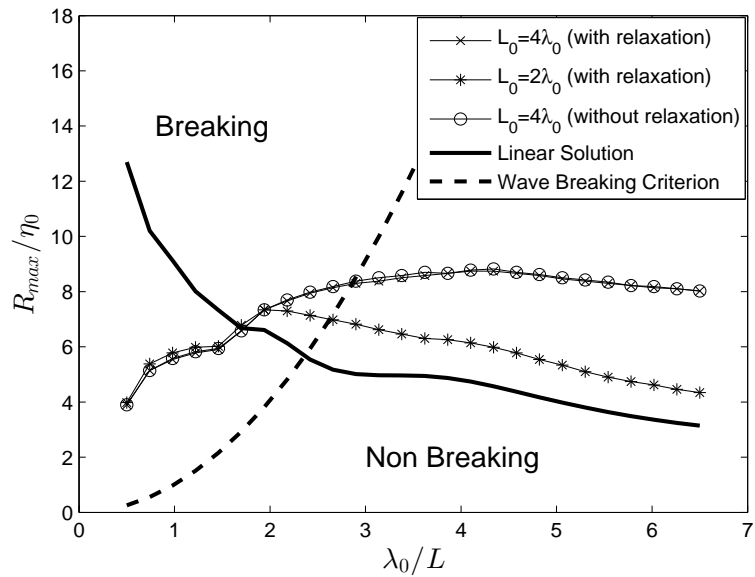}
   \end{center}
   \caption{\em Maximum run-up amplification as a function of nondimensional wavelength for the canonical case ($\eta_0 = 2.5$ \m, $h_0 = 100$ \m, $\tan \theta = 0.02$).}
   \label{fig:Amp25}
\end{figure}

In Fig.~\ref{fig:Steep} we see that the influence of $L_0$ is important and hence the wavefront steepness is critical to the run-up amplification. The existence of the relaxation zone does not affect the results when the constant depth region has a fixed length. The longer $L_0$ and therefore the wavefront steepness, the higher the run-up amplification, which in this case differs significantly from the theoretical curve (Fig.~\ref{fig:Amp25}), which is calculated for \emph{symmetric} monochromatic waves.

\begin{figure}
   \begin{center}
   \includegraphics[width=0.7\textwidth]{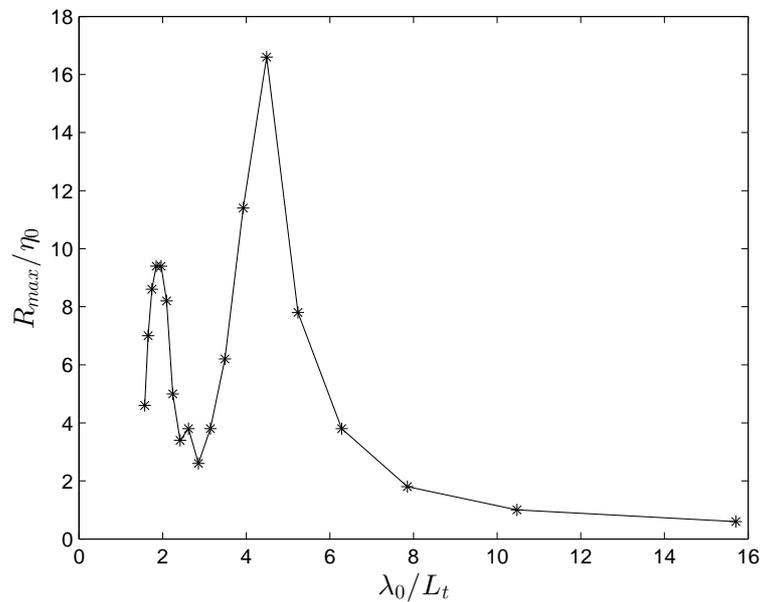}
   \end{center}
   \caption{\em Maximum run-up amplification as a function of nondimensional wavelength for the canonical case when $L_t = 8000$ \m is the distance from the undisturbed shoreline to the seaward boundary ($\eta_0 = 0.1$ \m, $h_0 = 100$ \m, $\tan \theta = 0.02$).}
   \label{fig:CanonicalRes}
\end{figure}

In the previous cases, we only considered waves that were shorter than the distance from the undisturbed shoreline to the seaward boundary. However, in the piecewise linear bathymetry (Fig.~\ref{fig:2slopes}), we found that resonance is possible for wavelengths larger than the distance mentioned above. The canonical case which we study in this section can be seen as the limiting example of the piecewise linear bathymetry as $\theta_i \rightarrow 0 \, ,\, i > 1$. Therefore, we decided to perform simulations using a plane beach with slope $\tan{\theta} = 0.02$ connected to a region of constant depth ($h_0 = 100$ m), which has a length $L_0 = 3000$ \m. This means that the distance from the initial shoreline to the seaward boundary is $L_t = 8000$ \m. We used very small amplitude waves ($\eta_0 / h_0 = 0.001$) and we did not put a relaxation zone close to the generation region. For each simulation we sent four non-breaking waves. We can observe in Fig.~\ref{fig:CanonicalRes} that resonance is possible for wavelengths larger than $L_t$ and this result is closer to our observations from the piecewise linear bathymetry.

\section{Discussion}

In this paper, based on the findings of \textsc{Stefanakis} \emph{et al.} (2011) \cite{Stefanakis2011}, we reproduced their results of run-up amplification using milder, more geophysically relevant bottom slopes and we showed that resonant run-up amplification on an infinite sloping beach is found for several waveforms and is robust to modal perturbations. 

Resonant run-up was confirmed by the laboratory experiments of \cite{Ezersky2012} for monochromatic waves and they also distinguished the resonant run-up frequencies from the natural frequencies of the system. The first resonant regime ($\lambda_0 / L = 5.2$, where $\lambda_0$ is the incoming wavelength and $L$ is the horizontal beach length) was achieved for non-breaking waves as in \cite{Ezersky2012}. Moreover, it is also interesting to note that our findings present similarities to those of \cite{Bruun1974, Bruun1977} who described a resonance phenomenon of short wave run-up on sloping structures. They described it as wave breaking taking place at the point of maximum run-down simultaneously with the arrival of the subsequent wave. Here, we do not see wave breaking, but there is a synchronization between the maximum run-down of a wave and the arrival of the next wave (Fig.~\ref{fig:SpatVel}). 

Run-up resonance in the laboratory experiments of \cite{Ezersky2012} was achieved for breaking waves as well, but for these cases they did not comment on the location where the breaking takes place. Hence, probably wave breaking is not the key factor to the resonant mechanism. Long wave breaking in the context of NSWE and its physical demonstration is a subtle issue. As it is noted by \cite{Synolakis1987}, the NSWE tend to predict wave breaking sooner than it actually happens in nature. Still there is an open question about whether tsunamis break when they shoal up a beach. \textsc{Madsen} \emph{et al.} (2008) \cite{MFS2008} suggest that the main flood wave does not break but instead short waves riding on top of the main tsunami do break, giving the impression that tsunamis break just before they reach the shoreline. 

The same resonant mechanism is found when the bathymetry is piecewise linear. However, when the beach is connected to a constant depth region, the picture is different. No resonant regimes are observed when the incoming wavelength is smaller than the distance between the initial shoreline and the seaward boundary. The maximum run-up amplification is found close to the breaking limit for nearly symmetric low amplitude waves. In that case the linear theory is in close agreement with the results for non-breaking waves. Nevertheless, the steepness of the wavefront plays an important role on run-up, with increasing steepness leading to higher run-up. It is not clear though if it is the wavefront steepness which is responsible for the increase of run-up values or the wave asymmetry (skewness). Increasing the incoming wavelength more than the wave propagation distance to the undisturbed shoreline results in observing resonant regimes similar to those found in the piecewise linear bathymetry example, which can be thought as the limiting case when the angles $\theta_i \rightarrow 0 \, ,\, i>1$.\footnote{Even though the length of the computational domain is not a physical parameter, it is of importance from an operational point of view, when one wants to predict run-up elevation based on recorded wave signals at an offshore location. In that case, the incoming wavelength can be either larger or smaller than the beach length.} It is of interest to report that \textsc{Keller} \& \textsc{Keller} (1964) \cite{Keller1964} tried to reproduce numerically their analytical solution and found a peak which corresponds to the resonant frequency in our simulations. However, they dismissed these results by saying that their computational scheme was not good enough. On the theoretical side, we can say that in linear theory, the existence or not of resonance depends on the geometry the bathymetry has at the seaward boundary. 

The discrepancies of the results using the two bathymetric profiles raise questions about the role boundary conditions play both physically and numerically, and more importantly about the character of the flow (stationary vs transient). The problem of long wave run-up has been attacked primarily from a stationary point of view in the past. The well-known solution \eqref{shore1} is a standing wave solution and as such does not exhibit any net propagation of energy over time. The solutions we investigated numerically are transient and as such can exhibit an amplification of energy over time.

\section*{Acknowledgements}
\addcontentsline{toc}{section}{Acknowledgments}

This work was funded by EDSP of ENS-Cachan, the Cultural Service of the French Embassy in Dublin (first author), the China Scholarship Council (second author), the ERC under the research project ERC-2011-AdG 290562-MULTIWAVE, SFI under the programme ERC Starter Grant - Top Up, Grant 12/ERC/E2227 and the Strategic and Major Initiatives scheme of University College Dublin.

\addcontentsline{toc}{section}{References}
\bibliographystyle{abbrv}
\bibliography{biblio}

\begin{thebibliography}{10}

\bibitem{Agnon1988}
Y.~Agnon and C.~C. Mei.
\newblock {Trapping and resonance of long shelf waves due to groups of short
  waves}.
\newblock {\em J. Fluid Mech}, 195:201--221, 1988.

\bibitem{Ahrens1981}
J.~P. Ahrens.
\newblock {Irregular wave runup on smooth slope}.
\newblock Technical report, CETA No. 81-17. U.S. Army Corps of Engineers,
  Coastal Engineering Research Center, FT. Belvoir, VA, 1981.

\bibitem{Antuono2007}
M.~Antuono and M.~Brocchini.
\newblock {The Boundary Value Problem for the Nonlinear Shallow Water
  Equations}.
\newblock {\em Stud. Appl. Math.}, 119:73--93, 2007.

\bibitem{Billingham2001}
J.~Billingham and A.~King.
\newblock {\em {Wave motion}}.
\newblock Cambridge University Press, Cambridge, 2001.

\bibitem{Brocchini1996}
M.~Brocchini and D.~H. Peregrine.
\newblock {Integral flow properties of the swash zone and averaging}.
\newblock {\em J. Fluid Mech}, 317:241--273, 1996.

\bibitem{Bruun1977}
P.~Bruun and A.~R. Gunb\"ak.
\newblock {Stability of sloping structures in relation to $\xi =
  \tan\alpha/\sqrt{H/L_0}$ risk criteria in design}.
\newblock {\em Coastal Engineering}, 1:287--322, 1977.

\bibitem{Bruun1974}
P.~Bruun and P.~Johannesson.
\newblock {A critical review of the hydraulics of rubble mound structures}.
\newblock Technical report, The Norwegian Institute of Technology Trondheim,
  1974.

\bibitem{Carrier1966}
G.~F. Carrier.
\newblock {Gravity waves on water of variable depth}.
\newblock {\em J. Fluid Mech}, 24(04):641--659, 1966.

\bibitem{CG58}
G.~F. Carrier and H.~P. Greenspan.
\newblock {Water waves of finite amplitude on a sloping beach}.
\newblock {\em J. Fluid Mech.}, 2:97--109, 1958.

\bibitem{CWY}
G.~F. Carrier, T.~T. Wu, and H.~Yeh.
\newblock {Tsunami run-up and draw-down on a plane beach}.
\newblock {\em J. Fluid Mech.}, 475:79--99, 2003.

\bibitem{Dutykh2011e}
D.~Dutykh, T.~Katsaounis, and D.~Mitsotakis.
\newblock {Finite volume schemes for dispersive wave propagation and runup}.
\newblock {\em J. Comput. Phys}, 230(8):3035--3061, Apr. 2011.

\bibitem{Dutykh2009a}
D.~Dutykh, R.~Poncet, and F.~Dias.
\newblock {The VOLNA code for the numerical modeling of tsunami waves:
  Generation, propagation and inundation}.
\newblock {\em Eur. J. Mech. B/Fluids}, 30(6):598--615, 2011.

\bibitem{Ezersky2012}
A.~Ezersky, N.~Abcha, and E.~Pelinovsky.
\newblock {Physical simulation of resonant wave run-up on a beach}.
\newblock {\em Nonlin. Processes Geophys.}, 20:35--40, July 2013.

\bibitem{Fritz2007}
H.~M. Fritz, W.~Kongko, A.~Moore, B.~McAdoo, J.~Goff, C.~Harbitz, B.~Uslu,
  N.~Kalligeris, D.~Suteja, K.~Kalsum, V.~V. Titov, A.~Gusman, H.~Latief,
  E.~Santoso, S.~Sujoko, D.~Djulkarnaen, H.~Sunendar, and C.~Synolakis.
\newblock {Extreme runup from the 17 July 2006 Java tsunami}.
\newblock {\em Geophys. Res. Lett.}, 34:L12602, 2007.

\bibitem{Fritz2011}
H.~M. Fritz, C.~M. Petroff, P.~A. Catal\'{a}n, R.~Cienfuegos, P.~Winckler,
  N.~Kalligeris, R.~Weiss, S.~E. Barrientos, G.~Meneses, C.~Valderas-Bermejo,
  C.~Ebeling, A.~Papadopoulos, M.~Contreras, R.~Almar, J.~C. Dominguez, and
  C.~E. Synolakis.
\newblock {Field Survey of the 27 February 2010 Chile Tsunami}.
\newblock {\em Pure Appl. Geophys.}, 168(11):1989--2010, Mar. 2011.

\bibitem{Ghidaglia2005}
J.-M. Ghidaglia and F.~Pascal.
\newblock {The normal flux method at the boundary for multidimensional finite
  volume approximations in CFD}.
\newblock {\em European Journal of Mechanics B/Fluids}, 24:1--17, 2005.

\bibitem{Grataloup2003}
G.~L. Grataloup and C.~C. Mei.
\newblock {Localization of harmonics generated in nonlinear shallow water
  waves}.
\newblock {\em Phys. Rev. E}, 68(2):26314, Aug. 2003.

\bibitem{Grilli2012}
S.~T. Grilli, J.~C. Harris, T.~S. {Tajalli Bakhsh}, T.~L. Masterlark,
  C.~Kyriakopoulos, J.~T. Kirby, and F.~Shi.
\newblock {Numerical Simulation of the 2011 Tohoku Tsunami Based on a New
  Transient FEM Co-seismic Source: Comparison to Far- and Near-Field
  Observations}.
\newblock {\em Pure Appl. Geophys.}, July 2012.

\bibitem{Kajiura1977}
K.~Kajiura.
\newblock {Local behaviour of tsunamis}.
\newblock In D.~Provis and R.~Radok, editors, {\em Waves on Water of Variable
  Depth}, volume~64 of {\em Lecture Notes in Physics}, pages 72--79. Springer
  Berlin / Heidelberg, 1977.

\bibitem{Kanoglu2006}
U.~Kanoglu and C.~Synolakis.
\newblock {Initial Value Problem Solution of Nonlinear Shallow Water-Wave
  Equations}.
\newblock {\em Phys. Rev. Lett.}, 97:148501, 2006.

\bibitem{Kanoglu1998}
U.~Kanoglu and C.~E. Synolakis.
\newblock {Long wave runup on piecewise linear topographies}.
\newblock {\em J. Fluid Mech.}, 374:1--28, 1998.

\bibitem{Kanoglu2013}
U.~Kanoglu, V.~V. Titov, B.~Aydin, C.~Moore, T.~S. Stefanakis, H.~Zhou,
  M.~Spillane, and C.~E. Synolakis.
\newblock {Focusing of long waves with finite crest over constant depth}.
\newblock {\em Proc. R. Soc. A}, 469(2153):20130015--20130015, Feb. 2013.

\bibitem{Keller1964}
J.~B. Keller and H.~B. Keller.
\newblock {Water wave run-up on a beach}.
\newblock Technical Report NONR-3828(00), Department of the Navy, Washington,
  DC, 1964.

\bibitem{Lamb1932}
H.~Lamb.
\newblock {\em {Hydrodynamics}}.
\newblock Cambridge University Press, 1932.

\bibitem{Lu2007}
Y.~Lu, H.~Liu, W.~Wu, and J.~Zhang.
\newblock {Numerical simulation of two-dimensional overtopping against seawalls
  armored with artificial units in regular waves}.
\newblock {\em Journal of Hydrodynamics, Ser. B}, 19(3):322--329, 2007.

\bibitem{Madsen2002}
P.~A. Madsen, H.~B. Bingham, and H.~Liu.
\newblock {A new Boussinesq method for fully nonlinear waves from shallow to
  deep water}.
\newblock {\em J. Fluid Mech.}, 462:1--30, 2002.

\bibitem{Madsen2008}
P.~A. Madsen and D.~R. Fuhrman.
\newblock {Run-up of tsunamis and long waves in terms of surf-similarity}.
\newblock {\em Coastal Engineering}, 55(3):209--223, Mar. 2008.

\bibitem{MFS2008}
P.~A. Madsen, D.~R. Fuhrman, and H.~A. Sch\"{a}ffer.
\newblock {On the solitary wave paradigm for tsunamis}.
\newblock {\em J. Geophysical Res.}, 113:C12012, 2008.

\bibitem{Mayer1998}
S.~Mayer, A.~Garapon, and L.~S. Sorensen.
\newblock {A fractional step method for unsteady free-surface flow with
  applications to non-linear wave dynamics}.
\newblock {\em Int. J. Num. Meth. Fluids}, 28(2):293--315, 1998.

\bibitem{Mei2005}
C.~C. Mei, M.~Stiassnie, and D.~K.-P. Yue.
\newblock {\em {Theory and applications of ocean surface waves: Linear
  aspects}}.
\newblock World Scientific, 2005.

\bibitem{Miles1971}
J.~W. Miles.
\newblock {Resonant response of harbours: an equivalent-circuit analysis}.
\newblock {\em J. Fluid Mech.}, 46(02):241--265, 1971.

\bibitem{Munk1964}
W.~Munk, F.~Snodgrass, and F.~Gilbert.
\newblock {Long waves on the continental shelf: an experiment to separate
  trapped and leaky modes}.
\newblock {\em J. Fluid Mech.}, 20(04):529--554, 1964.

\bibitem{Pelinovsky1992}
E.~N. Pelinovsky and R.~K. Mazova.
\newblock {Exact analytical solutions of nonlinear problems of tsunami wave
  run-up on slopes with different profiles}.
\newblock {\em Nat. Hazards}, 6(3):227--249, 1992.

\bibitem{Rabinovich1992}
A.~B. Rabinovich and A.~S. Leviant.
\newblock {Influence of Seiche Oscillations on the Formation of the Long-Wave
  Spectrum Near the Coast of the Southern Kuriles}.
\newblock {\em Oceanology}, 32:17--23, 1992.

\bibitem{Rabinovich2009}
A.~B. Rabinovich, I.~Vilibic, and S.~Tinti.
\newblock {Meteorological tsunamis: Atmospherically induced destructive ocean
  waves in the tsunami frequency band}.
\newblock {\em Phys. Chem. Earth. (B)}, 34(17-18):891--893, 2009.

\bibitem{shi2012}
F.~Shi, J.~T. Kirby, J.~C. Harris, J.~D. Geiman, and S.~T. Grilli.
\newblock {A high-order adaptive time-stepping TVD solver for Boussinesq
  modeling of breaking waves and coastal inundation}.
\newblock {\em Ocean Modelling}, 43-44:36--51, 2012.

\bibitem{Stefanakis2011}
T.~Stefanakis, F.~Dias, and D.~Dutykh.
\newblock {Local Runup Amplification by Resonant Wave Interactions}.
\newblock {\em Phys. Rev. Lett.}, 107:124502, 2011.

\bibitem{Synolakis1987}
C.~Synolakis.
\newblock {The runup of solitary waves}.
\newblock {\em J. Fluid Mech.}, 185:523--545, 1987.

\bibitem{noaa_report}
C.~E. Synolakis, E.~N. Bernard, V.~V. Titov, U.~Kanoglu, and F.~I. Gonzalez.
\newblock {Standards, criteria, and procedures for NOAA evaluation of tsunami
  numerical models}.
\newblock Technical report, NOAA/Pacific Marine Environmental Laboratory, 2007.

\bibitem{Synolakis1993}
C.~E. Synolakis and J.~E. Skjelbreia.
\newblock {Evolution of Maximum Amplitude of Solitary Waves on Plane Beaches}.
\newblock {\em J. Waterway, Port, Coastal and Ocean Engineering},
  119(3):323--342, May 1993.

\bibitem{TS94}
S.~Tadepalli and C.~E. Synolakis.
\newblock {The run-up of N-waves on sloping beaches}.
\newblock {\em Proc. R. Soc. Lond. A}, 445:99--112, 1994.

\bibitem{Tadepalli1996}
S.~Tadepalli and C.~E. Synolakis.
\newblock {Model for the leading waves of tsunamis}.
\newblock {\em Phys. Rev. Lett.}, 77:2141--2144, 1996.

\bibitem{Tissier2012}
M.~Tissier, P.~Bonneton, F.~Marche, F.~Chazel, and D.~Lannes.
\newblock {A new approach to handle wave breaking in fully non-linear
  Boussinesq models}.
\newblock {\em Coastal Engineering}, 67:54--66, 2012.

\bibitem{Tsuji1995}
Y.~Tsuji, F.~Imamura, H.~Matsutomi, C.~E. Synolakis, P.~T. Nanang, Jumadi,
  S.~Harada, S.~S. Han, K.~Arai, and B.~Cook.
\newblock {Field survey of the East Java earthquake and tsunami of June 3,
  1994}.
\newblock {\em Pure Appl. Geophys.}, 144:839--854, 1995.

\end{thebibliography}

\end{document}